# Bayesian Reasoning and Evidence Communication

Steven Lund (steven.lund@nist.gov) and Hari Iyer (hari@nist.gov)
Statistical Engineering Division, ITL/NIST, Gaithersburg, MD 20899
Contact author email: Steven.Lund@nist.gov

## Abstract

Many resources for forensic scholars and practitioners, such as journal articles, guidance documents, and textbooks, address how to make a value of evidence assessment in the form of a likelihood ratio ($LR$) when deciding between two competing propositions. These texts often describe experts presenting their $LR$ values to other parties in the judicial system, such as lawyers, judges, and potentially jurors, but few texts explicitly address how a recipient is expected to utilize the provided $LR$ value. Those that do often imply, or directly suggest, a hybrid modification of Bayes' rule in which a decision maker multiplies their prior odds with another person's assessment of $LR$ to obtain their posterior odds. In this paper, we illustrate how someone adhering to Bayesian reasoning would update their personal uncertainty in response to someone else presenting a personal $LR$ value (or any other form of an opinion) and emphasize that the hybrid approach is a departure from Bayesian reasoning. We further consider implications of recipients adhering to Bayesian reasoning on the role and ideal content of expert's reports and testimony and address published responses to our 2017 paper (Lund and Iyer, 2017), where we previously argued that the hybrid equation is not supported by Bayesian reasoning.

**Keywords:** Likelihood Ratios, Strength of Evidence, Weight of Evidence, Bayes Rule, Expert Opinions.

# 1 Introduction

Bayesian reasoning refers to the practice of evaluating and updating uncertainties in a manner that conforms with Bayes' rule, which describes the constraints the laws of probability place on how an individual's beliefs should be affected by new information, say $E$. More specifically, it shows how three probabilistic quantities, namely prior odds, likelihood ratios (or Bayes' factors), and posterior odds, must be related to one another in order to comply with the laws of probability. In its simplest application, Bayes' rule would apply to a person updating their belief regarding which of two propositions, say $H_1$ and $H_2$, is true in light of some newly considered information, say $E$. For instance, $H_1$ may reflect a perspective corresponding to the prosecution, and $H_2$ may reflect a perspective corresponding to the defense[1].

Suppose an individual characterizes how sure they feel about the truth of $H_1$ using probability $p_1$ and about the truth of $H_2$ using probability $p_2$. The ratio of these two probabilities, $O_{12} = \dfrac{p_1}{p_2}$, is often called the odds of $H_1$ versus $H_2$.[2] Upon encountering new information, say $E$, the individual's uncertainties may change. Thus, there is a distinction between the individual's probabilities as assessed before, or prior to, learning $E$ and the individual's probabilities as assessed after, or posterior to, learning $E$. The individual's original probabilities, $p_1$ and $p_2$, evaluated before encountering the new information, are referred to as prior probabilities (relative to $E$) and their ratio as prior odds of $H_1$ versus $H_2$. Probabilities reflecting the individual's uncertainty in the truth of $H_1$ and $H_2$ after learning the new information, say $p_1^*$ and $p_2^*$, respectively, are referred to as posterior probabilities (relative to $E$), and the ratio $O_{12}^* = \dfrac{p_1^*}{p_2^*}$ is referred to as posterior odds of $H_1$ versus $H_2$. The impact of the new information $E$ on the individual's uncertainty is reflected by the ratio $\frac{O_{12}^*}{O_{12}}$. For instance, if the prior and posterior odds of $H_1$ versus $H_2$ are very similar, their ratio will be close to 1, indicating that the new information has had very little impact. Bayes' rule relates this ratio of posterior and prior odds of $H_1$ versus $H_2$ to what is known as a likelihood ratio, the ratio between how likely one feels $E$ would be to occur if $H_1$ were true, say $l_1$, and how likely one feels $E$ would be to occur if $H_2$ were true, say $l_2$. In particular, Bayes' rule requires that an individual's ratio of odds, $\frac{O_{12}^*}{O_{12}}$, is equal to the individual's likelihood ratio, $LR_{12} = \frac{l_1}{l_2}$. This can be equivalently restated as a requirement that an individual's posterior odds are equal to the product of the individual's corresponding prior odds and the individual's likelihood ratio. Any triplet of prior odds, likelihood ratio, and posterior odds that fail to conform to Bayes rule violates the basic axioms of probability theory and may be labeled as irrational or incoherent.[3]

---

[1] In order for someone to arrive at updated or "posterior" probabilities of the propositions given the presented evidence using Bayes' rule, the set of propositions considered must be exhaustive from their perspective. That is, any proposition for which the individual provides a non-zero prior probability must be included in the set of considered propositions. The use of exhaustive proposition sets is a fundamental element of Bayesian reasoning that has sometimes been modified in its use in forensic applications.

[2] By formal convention, the odds for $H_1$ refers to the ratio $O_1 = \dfrac{p_1}{1 - p_1}$, which is equivalent to $O_{12}$ if and only if $H_1$ and $H_2$ are mutually exclusive and exhaustive so that $p_1 + p_2 = 1$.

Thus, Bayes' rule provides a pathway for how a rational person could update their uncertainty, after encountering new information, in three steps: (1) Assess how sure you feel about the truth of each considered proposition; (2) Assess how likely the newly encountered information would be to have occurred, assuming the truth of each considered proposition in turn; (3) Derive posterior probabilities in accordance with Bayes' rule.[4] This general process is the same, regardless of what form the new information takes.

Throughout this paper, we specify who has assigned the value of each probability. While somewhat tedious, we do this to consistently acknowledge that there are multiple individuals who consider forensic evidence and that probabilities are personal.[5] While rules of probability may seem rigorous and exact, they are "if-then" statements as in, "if your prior odds are 0.1 and your likelihood ratio is 100, then your posterior odds should be 10." The laws of probability do not tell you what priors to start with or what likelihoods to use, so they do not dictate the appropriate probabilistic interpretation of any given situation. Consequently, probabilistic assessments can be expected to vary from one person to another, so it is important to specify whose probabilities are being discussed in real world applications involving multiple people processing information and assigning probabilities. Lindley's (2013) use of the word "your" emphasizes this point in writing, "We saw in §6.6 how evidence $E$ before the court would change your probability to $p(G|EK)$ using Bayes' rule. The calculation required by the rule needs your likelihood ratio $p(E|GK)/p(E|G^cK)$, involving your probabilities of the evidence, both under the supposition of guilt and of innocence..."

---

[3] When there are more than two mutually exclusive propositions, Bayes' rule can still be used to obtain posterior probabilities as a function of prior probabilities and likelihoods associated with each proposition. For ease of discussion and without loss of generality, we restrict our presentations to the simplest scenario of two mutually exclusive propositions.

[4] Note that a person can also choose to wait to conduct an evaluation of uncertainty until after having received all information relevant to their decision and proceed directly to specifying "posterior" odds of $H_1$ versus $H_2$ without specifying any other priors or likelihood ratios. If the individual did happen to articulate what their odds of $H_1$ versus $H_2$ would have been without a given piece of information and a likelihood ratio for that piece of information, then that triplet of prior odds, likelihood ratio, and posterior odds would need to satisfy Bayes' rule to be considered rational and coherent.

Many people believe that the role of forensic experts is to communicate a likelihood ratio as their opinion of the value of the evidence they have considered with respect to a pair of propositions, $H_1$ and $H_2$. Given that probabilities are personal, the question arises, "what is a recipient supposed to do with an expert's $LR$?" Though it may seem unintuitive, learning the value of an expert's $LR$ is a new piece of information for a recipient and can be processed by the recipient to form their own $LR$ for the expert's $LR$. De Finetti (2017) describes this challenge as "subjective squared: our subjective judgment regarding the subjective judgment of others." We illustrate this process in Section 3 by applying Bayesian reasoning within a simple model of evidence communication. The recipient's $LR$ for the expert's interpretation can be combined with the recipient's prior odds via Bayes' equation just like it could for any other piece of information. Few scholars or forensic experts advocate for this process,

---

[5]We consider the viewpoints expressed in this paper to be natural consequences of accepting that uncertainty is personal, which has been explicitly emphasized by scholars in probability and statistics (e.g., Lindley, 2013; Kadane, 2020; de Finetti, 2017) and in forensics (e.g., Taroni *et al.*, 2016; Berger and Slooten, 2016; Gittelson *et al.*, 2018). For the convenience of readers who may not be comfortable or familiar with the perspective that probabilities are personal, we also provide a brief discussion and an illustrative example in Appendix-A.

or even explicitly acknowledge it, in discussions or writings about Bayes' rule in forensic applications.[6]

Instead, when Bayes' rule is brought up in forensic contexts, it is often presented in general terms that do not specify to whom the probabilities belong. That is, an examiner may summarize their findings in the form of an $LR$ and state that Bayes' rule dictates that posterior odds are given by the product of prior odds with the $LR$. Some forensic science scholars, (for instance, Aitken and Nordgaard, 2018), explicitly represent recipients of an expert's $LR$ as combining their own prior odds with the expert's likelihood ratio to obtain their posterior odds in favor of $H_1$ compared to $H_2$.

In an earlier publication (Lund and Iyer, 2017 - hereafter referred to as `LI`), we pointed out that Bayes' rule only applies at an individual level and that this hybrid application is a departure from Bayesian reasoning. More explicitly, Bayes' rule as it applies to the decision maker ($DM$, e.g., a trier of fact) is

$$\text{Posterior Odds}_{DM} = \text{Likelihood Ratio}_{DM} \times \text{Prior Odds}_{DM}$$

whereas Bayes' rule as it applies to the expert is

$$\text{Posterior Odds}_{Expert} = \text{Likelihood Ratio}_{Expert} \times \text{Prior Odds}_{Expert}$$

A decision maker's substitution of the expert's $LR$ for their own would lead to a hybrid modification of Bayes' rule given by

$$\text{Posterior Odds}_{DM} = \text{Likelihood Ratio}_{Expert} \times \text{Prior Odds}_{DM}.$$

---

[6] However, the problem of updating one's own uncertainty about an event given the probability assessment by someone else for that event has been considered by many Bayesian experts; see, for instance Lindley *et al.* (1979), French (1980), Aspinall and Cooke (2013) and Ouchi (2004).

Although we are not aware of anyone explicitly arguing that decision makers should not use their own personal $LR$ to arrive at their posterior odds, we feel many presentations and discussions of weight of forensic evidence assessments indirectly promote the hybrid equation by omission. This occurs when someone presents a generic Bayes' equation and indicates that an expert will provide a (or especially "the") $LR$, without making it clear that each party has their own Bayes' equation. When recipients (especially those previously unfamiliar with Bayesian reasoning) are informed by a forensic expert that the appropriate interpretation of evidence is to multiply the recipient's prior odds with the likelihood ratio to get their posterior odds, and that the role of the forensic expert is to assess the likelihood ratio for the evidence and provide it to the recipient, it seems highly unlikely that the recipients would instinctively understand that the proper application of Bayesian reasoning is for them to form their own likelihood ratio for the information provided by the expert. This would be analogous to telling someone they should have coffee, telling them your job is to make coffee, bringing them a cup of coffee, and expecting that they will react by going make their own cup of coffee.

Here we provide some examples we view as indirectly promoting the hybrid Bayes' rule by omission, with added italics to highlight the most relevant phrases:

- "Bayes Theorem shows us that, while the investigator or court is concerned with questions of the type: 'what is the probability that the suspect was at the crime scene?', *the scientist, through the likelihood ratio, should address questions of the type 'what is the probability of the evidence given that the suspect was not at the crime scene?'*" (Evett, 1987)

- "The formula can be expressed in words as follows: Posterior odds = Likelihood ratio × Prior odds. The court is concerned with questions of the kind 'what is the probability that the defendant committed the crime given the evidence?' but *Bayes theorem demonstrates that, for the scientist to assist the court in updating its probabilities s/he* *must address questions of the kind 'what is the probability of the evidence given that the defendant committed the crime?'"* (Evett, 1998)

- "Bayes' Theorem provides a model that clearly distinguishes the role of the scientist and that of the fact finders. *The role of the scientist is to advise the fact finders on the strength of the evidence by assigning the LR. Any consideration of the prior or posterior odds (or the probability) of the propositions is left to the fact finders.*" (Buckleton *et al.*, 2020)

- "*The role of the forensic scientist is to assign the probabilities of the evidence given the propositions that are considered.*" (Buckleton *et al.*, 2020)

While the authors listed above may not have intended to support a hybrid Bayes' rule, the quotes above seem likely to suggest its use and make no mention of a recipient processing an expert's likelihood ratio to form their own personal weights of evidence. Other authors have appeared to endorse the application of the hybrid modification to Bayes' rule in evidence communication more explicitly. For instance,

- "Bayes Rule tells us that we then take those prior odds and multiply them by the likelihood ratio of the blood/DNA evidence in order to arrive at the posterior odds in favour of the defendant's paternity. The Court then has to consider whether those odds meet the required standard of proof. Thus the expert should say 'however likely you think it is that the defendant is the father on the basis of the other evidence, my evidence multiplies the odds X times'." (Robertson and Vignaux, 1992)

- "The main focus of attention will be confined to the perspective of how one can assess the value of scientific findings in order to inform about how findings should affect the views of others on selected issues in a case." (p. 182) (Biedermann, Taroni, *et al.*, 2014)

- "For example, for a likelihood ratio of a thousand, the scientist may think of reporting along the following lines: 'My findings are on the order of one thousand times more probable if the person of interest is the author of the questioned text than if an unknown person wrote the questioned text. Hence, whatever odds the recipient of expert information assesses that the person of interest is the author, based on other evidence, my findings multiply those odds by one thousand. For example, if the prior odds are even, then the posterior odds are one thousand, but will be less for smaller prior odds." (Biedermann, Bozza, *et al.*, 2018)

- "Determination of the BF (Bayes' factor) is typically considered to be in the domain of the forensic scientist." (Taroni *et al.*, 2016)

These viewpoints espouse precisely the application of the hybrid form of Bayes' rule to evidence communication, in which an expert is expected to provide the value of a likelihood ratio (or Bayes' factor) that someone else "should" use when applying Bayes' rule.

We believe that support for the hybrid Bayes' rule may result from failing to appropriately distinguish personal decision making from evidence communication. Communication, which concerns the transfer of information between two or more parties, and Bayesian reasoning, which concerns an individual reflecting on their own personal uncertainties, naturally occur in different contexts. Conflating communication and Bayesian reasoning can blur the roles of experts and the people they inform, thus leading to misapplications of Bayesian reasoning.

Expert reporting and testimony are communication processes. The purpose of these processes is for forensic experts to provide information to other members of the judicial system to help them make better decisions. Much attention has been given to what type of information experts should include in their communications. See, for instance, Blastland *et al.* (2020). Few of these writings consider formal representations of communication, which require, at a minimum, components for a sender, the message sent by the sender, a receiver, and the message received by the receiver. See, for instance, Figure 1 in Shannon's paper "A Mathematical Theory of Communication" (Shannon, 1948).

In a simple model of forensic evidence communication, the sender is a forensic expert, the message sent is the report or testimony issued from the expert, the receiver is another stakeholder in the judicial system, such as a lawyer, juror, or judge, and the message received is the stakeholder's understanding of the expert's report or testimony. In more realistic settings, there can be multiple experts, each of whom provide reports or testimony addressing the same pieces of evidence relative to the same pair of propositions, but arrive at different opinions. During expert testimony, recipients can also be exposed to information beyond what is reflected in an expert's opinion, during direct examination of witnesses as well as their cross examination. We believe the recipient's understanding of the provided message, which is not present in the hybrid modification to Bayes' rule, is an important component of assessing communication practices.

In the context of the simple evidence communication model, there are two parties that might apply Bayesian reasoning. Experts can think about Bayes' rule when formulating their assessments and could include their own personal probabilistic interpretations as part of the message sent in reports or testimony. Receivers may also apply Bayesian reasoning, either formally or informally, when processing their understandings of the messages received from experts. Accepting that probabilities are personal means acknowledging the expert and the recipient are two separate entities who cannot be expected to have the same probabilistic assessments, even in the scenario where both are envisioned as applying Bayesian reasoning and both have access to the same collection of empirical data.

Considering non-expert stakeholders in the judicial system as practicing Bayesians provides a clear separation between the personal uncertainties of the expert, the information the expert provides to a recipient, and the recipients' processing of the information they receive. Importantly, a recipient applying Bayesian reasoning would not directly adopt sentiments expressed by the experts (though their application of Bayesian reasoning could lead them to the same sentiment coincidentally). That is, they would not interpret an expert's opinion as an answer key that informs how they (the recipient) should react to the evidence. Instead, they would view the expert's opinion as a new piece of information and subsequently evaluate how the expert's opinion will influence their own beliefs by assessing how likely the expressed message would be to occur under each of the propositions of interest to them.

So how can an expert best inform a Bayesian recipient's interpretation of the output (i.e., result or opinion) provided in a given case? We believe that this remains an important, perhaps central, question when pursuing effective and scientifically justifiable expert communication practices. We further argue that considering this question from the perspective of applying the principles of Bayesian reasoning to evidence communication does not lead to a suggested practice of experts providing their personal $LRs$. Rather, it ultimately points to the importance of gathering and openly providing data regarding the performance of the analytical pipeline that leads to an output (e.g., black box or validation studies) and in-depth discussions of modeling uncertainty (uncertainty associated with selecting one of a collection of plausible statistical models to fit the empirical data). The role of these components are suppressed by representations of evidence communication in which recipients directly accept an expert's opinion as their own. Within this deferential model, recipients simply and completely "trust the expert," accepting as fact the subjective modeling choices that fall outside an expert's domain of expertise. These (mis)representations of Bayesian reasoning are devoid of the necessary and natural step in which recipients critically reflect upon information provided by the expert.

Acknowledging the step in which recipients critically reflect on offered interpretations or opinions naturally leads to questions regarding the performance of the methods applied by the expert. Bayes' rule suggests a rational recipient would evaluate their own personal likelihood of the offered opinion or result under each of the propositions of interest (to the recipient). That is, the rational recipient would consider questions of the form "how likely is the expert to have arrived at the offered result if proposition A were true?" The recipient would consider that general question, plugging in for A each proposition of interest in turn. Whether or not these questions are openly recognized by experts, the rational recipient would filter offered results or opinions through their own personal uncertainty regarding how the expert performs in cases like the one at hand. This step could lead to a substantial difference between the sentiment expressed by the expert and the interpretation of that sentiment by the recipient. In particular, large uncertainty regarding what results an expert tends to arrive at under different scenarios would likely severely limit the perceived value of even the strongest expressed opinions. This impartial and rational response may outwardly appear as skepticism. Of course, even in the absence of empirical performance assessments, a recipient could feel strongly confident that forensic methods are essentially foolproof or that the experts always know best (sometimes referred to as the "Reverse CSI effect," as in Godsey and Alou, 2010[7]). For recipients exhibiting the "Reverse CSI effect," their uncertainty could lead them to substantially amplify the strength of a moderate opinion (e.g., mapping an $LR$ of 2 to strongly suggesting guilt of the defendant because it is greater than 1). An expert could hope to strengthen the basis of a recipient's decision by readily communicating additional (factual) information that addresses the performance of their chosen method in cases like the one at hand. The motivation is not to bolster or justify the expert's offered opinion, but to enable recipients to reach their own informed interpretation of what the expert's opinion means to them. In other words, if Bayesian reasoning is adopted in the context of evidence communication, its proper place is with the recipient. Advocates of Bayesian reasoning should reject communication practices that suggest or imply the expert's role is to instruct or recommend to recipients a particular interpretation of evidence, in favor of practices that prompt recipients to form their own interpretation while providing understandable, accurate, and useful[8] information to aid this effort. While the specifics

---

[7]See also https://en.wikipedia.org/wiki/CSI_effect

[8]We find these attributes to be a more appropriate rubric for evaluating expert communication practices

will vary from discipline to discipline or even case to case, a general target is to provide or adequately summarize[9] available data that inform the distribution of outcomes the expert's approach produces in cases like the one at hand.

In some cases, an expert might expect that recipients will have sufficient experience or intuition regarding the critical elements of the evidence evaluation such that the recipients' priors regarding the distribution of provided results or opinions under the propositions of interest would not change substantially with the presentation of performance data. An example of this could be seeing a Nike brand logo in a crime scene footwear impression and an Adidas brand logo in the test impression. The more technical and specialized a form of evidence evaluation becomes, the more Bayesian recipients may need to rely on provided performance data, rather than underlying technical or scientific principles, to shape their likelihoods of the offered opinions. For instance, a recipient may not understand the science behind laser-ablation-inductively-coupled plasma-mass spectrometry (LA-ICP-MS) well enough to have a good sense of how often experts using the methodology might mistake glass shards from one manufacturer as having come from a different manufacturer. However, even a recipient who does not know what LA-IMP-MS stands for could understand information of the sort that experts using LA-IMP-MS made no such mistakes in 100 blind tests in which they received glass shards from one manufacturer and were asked to compare them to glass from another manufacturer.

Careful separation of Bayesian reasoning from evidence communication also clarifies the types of information the recipient can effectively use. The interpretive step in which recipients consider how likely the expert is to have expressed the provided opinion under each

---

than "rational" or "coherent." While an expert's handling of uncertainty when arriving at a *LR* may strictly abide by the rules of probability, the communication practice of expert's providing *LR*s does not seem rational, especially when it is widely accepted that a substantial proportion of potential recipients do not understand them.

[9]Note that these descriptions must include specific results obtained from using the method. High level summaries of the amount of training or experience an expert has in applying the method, the number of peer-reviewed publications associated with the method, or a statement that the method has been "validated" do not carry specific implications regarding what is known about the method's performance.

of the propositions of interest to them (the recipient) strips the offered opinion of its literal meaning. This makes Bayesian recipients rather indifferent to the form of opinion an expert uses to summarize their interpretation of the evidence in the case at hand. In particular, they can work effectively with numeric $LR$s or their verbal counterparts, categorical conclusions, or even arbitrary units. The interpretive step also acts as an internal calibration, mapping from whatever scale the expert originally used to the scale that is consistent with the recipients' own uncertainties. From this perspective, an expert presenting evidence to a Bayesian recipient should express opinions using whatever form has the strongest demonstrated performance.

To be clear, we are not decrying the use of expert opinions. Experts play a critical role in identifying, collecting, and analyzing evidence. Without the skills of forensic experts, other members of the judicial system would have to attempt to deal with complex and chaotic crime scenes and highly technical applications of chemistry, physics, and biology that require years of training and experience to perform correctly. They would have to guess when sifting out subtle discriminating features and comparing complex patterns awash with a sea of variability and interference. No one is suggesting that other members of the judicial system are capable of performing the analysis of evidence themselves. Recipients, Bayesian or otherwise, trust and depend on trained experts as being the most capable of transforming raw information in the form of physical evidence into some form of ranking system that offers the optimal discrimination among any methods discovered to date. Oftentimes, the most effective methods for discriminating between $H_1$ and $H_2$ (e.g., whether or not two footwear impressions were made by the same shoe) we have to date are the examiner's experience-based opinions themselves. Our discussions in this paper are not intended to discourage the use of examiner opinions or judgement, but to reflect how evidence communication can best support examiner opinions in a logical and scientifically sound manner. To that end, we are advocating for factual reporting and testimony about method performance, regardless of the extent to which that method depends on expert opinions and judgements.

We present a representation of the evidence communication process in Section 2. In Section 3, we explore how a Bayesian fact finder would apply Bayesian reasoning in processing the information communicated by an expert and illustrate the potential benefits of making available performance data related to expert opinions. In Section 4, we examine the task faced by a non-Bayesian recipient of expert opinion. We note there that a lay person fact finder may not be aware of the limits of the domains of expertise of the expert and thus may be inclined to take the expert's opinion at face value. With this in mind, we explore, in Section 5, what the limits of expertise are for a well-trained expert statistician since statisticians, or those trained in the methods of probability and statistics, are the experts who provide probabilistic opinions or develop algorithms or methods that output probabilities for use by forensic practitioners. We conclude the article with some summary statements in Section 6. Appendix A expounds on the notion that probability is personal for readers not familiar with this viewpoint. Appendix B addresses published responses to `LI`.

## 2  Representing the Evidence Communication Process

In this section we outline a simple framework to provide context for properly understanding the roles of experts and recipients in evidence communication and how Bayesian reasoning relates to this framework. Figure 1 provides a starting point. The activities and outputs related to the expert are shown in blue and the activities and outputs related to the recipient (who could be thought of as a trier of fact, for concreteness, although there are other recipients such as the defense attorney, the defendant, the prosecutor, etc., who all have to make some choices before the case ever goes to trial) are shown in green.

The expert (B) receives or discovers evidential material $E$ to be processed and considers two or more propositions based on input from requesting parties. The forensic expert will conduct an analysis chosen based on their training, experience, protocols, or intuition and the considered propositions. This may involve an initial assessment regarding the merits of further analysis, which could be applied following practices of Case Analysis and Interpretation (Cook *et al.*, 1998). Let the examiner's understanding of the evidence following analysis be denoted as $\phi_B(E_B, I_B)$. Here $I_B$ represents background information available to $B$ including training, experience, standard operating procedures, perceived incentives, factual background information, modeling choices, etc.[10]

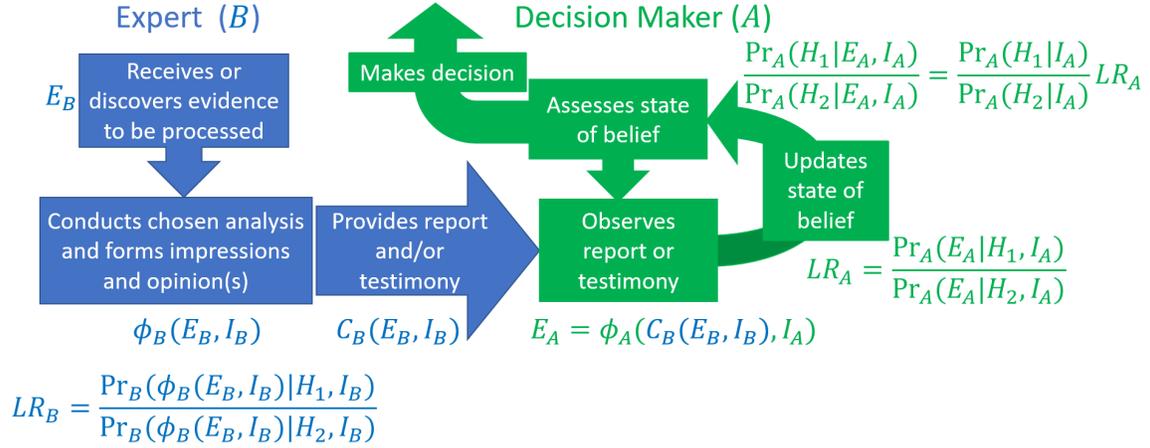

Figure 1: Bayesian reasoning applied within simple evidence communication model

Following these endeavors, the forensic examiner will write a report, verbally testify, or otherwise present their evaluation of the evidence. This communication is unlikely to be the original physical evidence itself, but a reflection of the insights an expert has gained by interacting with the physical evidence during the course of analysis. The utility of considering the communication rather than original physical evidence comes from the transformation of raw physical information, which generally requires advanced training to make sense of, into a simpler output that recipients can understand without specialized knowledge and expertise.

As such, let us denote what is presented by the forensic examiner as $\mathcal{C}_B(E_B, I_B)$ ($\mathcal{C}$ stands for 'communication' and the subscript $B$ stands for 'by $B$'). This presentation is a function of the evidence itself (as understood by $B$), $E_B$, and other factors that affect how the examiner interacts with, perceives, and, ultimately, characterizes the evidence. These other factors are also summarized within $I_B$.

---

[10] Some components of $I_B$ could be characterized as factual information that is not disputed by others. However, $I_B$ also includes components, for instance modeling choices, that others may not accept as factual. In essence, $I_B$ consists of everything that $B$ *believes* to be true. It is our view that $I_B$ should not be described as *background information* since it gives the misleading impression that all components of $I_B$ are factual.

Despite the best of intentions, recipients will likely pay imperfect attention to, or have imperfect comprehension and/or retention of, the forensic examiner's presentation. These factors contribute to the noise term in Shannon's communication model and distort what is received by the recipient into something different than what was expressed by the expert.[11] This potentially distorted message is of primary interest because it is the only function of $E_B$ the recipient has to work with. Let $E_A = \phi_A(\mathcal{C}_B(E_B, I_B), I_A)$ represent what the recipient retains regarding the expert's presentation, report, or testimony. Here $I_A$ represents any background information possessed and used by the decision maker ($A$), as well as any propositions or events that $A$ believes to be true, to form their uncertainty assessments or to make decisions. If more than one expert provides a report or testimony about a given component of evidence, the recipients' understanding can be extended to be a function of the communications from each of $N$ presenting experts as $E_A = \phi_A(\{\mathcal{C}_{B_i}(E_B, I_{B_i})\}_{i=1}^N, I_A)$, where $B_i$ is an index for expert $i$.[12] After receiving the communications from the expert(s), recipients will consider the newly acquired information, update their beliefs, and ultimately make their decisions.

In this simple evidence communication model, there are two parties that could be viewed as applying Bayesian reasoning. Experts could think about Bayes' rule when formulating their assessments and could choose to include their personal probabilistic interpretations as part of the message provided in reports or testimony. Receivers could also apply Bayesian reasoning when processing their understandings of the messages received from experts. Even if both the expert and the recipient are envisioned to apply Bayesian reasoning, the expert and the recipient remain two separate entities. Both parties applying Bayesian reasoning does not mean they will come to share one mind or opinion. Accordingly, any theoretical assessment of communication practices should clearly reflect the fact that uncertainty is personal (Lindley, 2013; Kadane, 2020; de Finetti, 2017; Biedermann, 2013; Berger and Slooten, 2016).

---

[11]"Transposing the conditional," when recipients confuse an expert's likelihood ratio to be the expert's posterior odds, is a commonly occurring example (Thompson *et al.*, 2013).

[12]We note that the hybrid modification of Bayes' equation does not accommodate multiple expert LRs for the same piece of evidence. Multiplying prior odds by two different LRs produces two different posterior odds. Obviously, the recipient would need to form their own LR based on the information provided by all experts. Our perspective is that this is the correct framework even when a given component of evidence is only addressed by a single expert.

In the following sections we illustrate the process of a recipient applying Bayes' rule in response to an expert's communication. We additionally consider a recipient who is unfamiliar with Bayes' rule.

## 3 Recipients Applying Bayesian Reasoning

Applying Bayesian reasoning generally[13] means someone updates their prior beliefs using their own likelihoods for the newly received information under each proposition of interest to them, no matter what form that information takes. Throughout this section we use the phrase "Bayesian recipient" as shorthand for a hypothetical recipient applying Bayesian reasoning.

In the simplest case where recipients consider exactly two propositions, Bayesian recipients would update their prior odds using their likelihood ratios for the newly received information, namely

$$LR_A = \frac{Pr[E_A|H_1, I_A]}{Pr[E_A|H_2, I_A]}. \tag{1}$$

---

[13]We say "generally" because one does not need to form or articulate prior probabilities and likelihoods to comply with Bayesian reasoning. One can decide to directly assess their personal (posterior) uncertainty after absorbing whatever information has been provided. See Lindley (2013) and Good (1991); See also footnote 4.

In situations where recipients are interested in more than two propositions, they would evaluate Bayes' factors, which depend on prior probabilities given to each proposition and the corresponding likelihoods for the newly received information, to update their posteriors.

Note that the steps to apply Bayesian reasoning do not depend on what type of information an expert provides. An expert's presentation of $\phi_B(E_B, I_B)$ could be factual or an opinion, with the latter being expressed as a categorical conclusion, a numeric likelihood ratio, a perceived level of similarity, etc. In any case, Bayesian recipients consider the likelihood of having received this new information, which we have denoted as $E_A$, under each proposition of interest to them. Critically, Bayesian recipients do not view offered opinions as the logical solution to how they should feel about the evidence or simply accept them at face value. Instead, they would consider others' opinions as new information and use their own critical thinking to assess what that new information means to them.

For an expert's message to substantially impact Bayesian recipients' posteriors for the propositions of interest, their likelihoods of encountering this information must differ greatly under the various propositions of interest. In circumstances where the recipient is unfamiliar with the methods the expert used, they may reflect their initial lack of knowledge or skepticism about method output by specifying "vague" or "non-informative" priors for the distribution of method outputs under each of the propositions of interest to them. In order for an expert's opinion to be strongly impactful, it must be accompanied by enough supporting information to transform the recipient's vague priors for the distributions of method outputs into posteriors that are highly differentiated across the propositions of interest to the recipient. Explaining what aspects of the current assessment led to the offered interpretation provides some insight to recipients, but this is often technical and specialized (meaning it may be familiar to experts but not to layperson recipients), which risks doing more to impress or confuse than to inform and empower the recipients. Technical details are only helpful to the extent that they reduce the Bayesian recipients' uncertainty regarding the distribution of outcomes under the various propositions they consider. If recipients do not comprehend the technical details or their implications, these descriptions may have little effect on their uncertainty regarding the distributions of outcomes.

Fortunately, there is another option that is simple and direct. Factual information describing what opinions or results have occurred, in ground truth known instances representing the propositions expected to be of interest to the recipients, provides a scientifically sound and easily understood basis to aid the recipients' assessments of these likelihoods. We illustrate this point with a simple example in the following section.

## 3.1 Example of Bayesian Recipient Receiving Expert Communication

Consider a burglary case where glass fragments were found on the clothing of a person of interest that are thought to have come from a broken window at the house where burglary occurred. The propositions considered by the forensic examiner are

$H_1$ : The fragments on the clothing came from the broken window at the house where the burglary occurred

$H_2$ : The fragments came from a different source

After a careful analysis of the evidence the forensic expert summarizes his/her findings in the form of a strength of evidence statistic expressed as a likelihood ratio, which we write as $LR_B$ to indicate that it is the strength of evidence as assessed by $B$, where

$$LR_B = \frac{Pr_B[E_B|H_1, I_B]}{Pr_B[E_B|H_2, I_B]}.$$

The notation $Pr_B$ refers to probability assessments made by $B$. The value of $LR_B$ is communicated to a recipient $A$ who understands it so that $E_A = LR_B$. Suppose the value of $LR_B$ is $r$ and that $H_1$ and $H_2$ are also the propositions of interest to $A$. Prior to receiving the $LR_B$ information, suppose $A$ represents her/his uncertainty regarding the truth of $H_1$ in the form of prior odds, which for this example, we take to be $\theta$ (known to $A$). Upon hearing $LR_B$, $A$ is interested in $Pr_A[H_1|LR_B = r, I_A]$, $A$'s updated belief regarding the truth of $H_1$. Using Bayes' rule to modify $\theta$ into $Pr_A[H_1|LR_B = r, I_A]$ requires $A$ to specify his/her beliefs regarding the probabilities that $LR_B = r$ under $H_1$ and also under $H_2$. Suppose that $A$ has not previously encountered validation data for the expert's chosen method of assessing an $LR$ and feels unsure of what type of values the $LR$ method would produce when applied to glass fragments from a common source (corresponding to $H_1$) or when applied to glass fragments from different sources (corresponding to $H_2$). This uncertainty is reflected by the prior $A$ chooses to represent the potential $LR_B$ distributions under $H_1$ and $H_2$. For the sake of illustration, suppose the recipient assumes that $log(LR_B)$ is normally distributed for $H_1$ and $H_2$, respectively, but is uncertain about the mean and variance for each of these two distributions. That is, the recipient assumes

$$log(LR_B)|H_1, \mu_1, \sigma_1^2, \mu_2, \sigma_2^2 \sim \text{Normal}(\mu_1, \sigma_1^2)$$
$$\text{and } log(LR_B)|H_2, \mu_1, \sigma_1^2, \mu_2, \sigma_2^2 \sim \text{Normal}(\mu_2, \sigma_2^2).$$

Suppose the recipient chooses to convey their uncertainty in the parameters $(\mu_1, \sigma_1^2)$ and $(\mu_2, \sigma_2^2)$ using normal-gamma distributions, which are the conjugate priors for normal distributions with unknown means and variances (page 268, Bernardo and Smith, 2009), meaning it is computationally simple to update these priors based on new information from examiner performance. To simplify notation, precision (i.e., 1 over variance) is used in place of variance. That is, we use $\tau_1 = 1/\sigma_1^2$ and $\tau_2 = 1/\sigma_2^2$.

In particular, suppose the recipient's prior for $(\mu_1, \tau_1)$, which is given according to $(\mu_1, \tau_1) \sim \text{Normal-Gamma}(\mu_{10}, n_{\mu 1}, \tau_{10}, n_{\tau 1})$, is specified in terms a prior mean of $\mu_{10} = 5$ with $n_{\mu 1} = 1$ observation's worth of information about the mean and a prior precision of $\tau_{10} =$ $\frac{1}{100}$ with $n_{\tau 1} = 1$ observation's worth of information about the precision. Similarly, the recipient's prior for $(\mu_2, \tau_2)$, which is given according to $(\mu_2, \tau_2) \sim$ Normal-Gamma$(\mu_{20}, n_{\mu 2}, \tau_{20}, n_{\tau 2})$, is specified as a prior mean of $\mu_{20} = -5$ with $n_{\mu 2} = 1$ observation's worth of information about the mean and a prior precision of $\tau_{20} = \frac{1}{100}$ with $n_{\tau 2} = 1$ observation's worth of information about the precision. Finally, suppose the recipient assumes the pair $(\mu_1, \tau_1)$ to be independent of the pair $(\mu_2, \tau_2)$. The corresponding marginal distributions for $LR_B$ under each proposition and the corresponding $LR_A$ values are shown in Figure 2.

As seen in Figure 2, for the priors $A$ has chosen, even extreme values of $LR_B$ will have little influence on $A$.[14] For the examiner's opinion to have the potential to make a substantial difference in $A$'s uncertainty regarding the truth of $H_1$, $A$ must also receive information that reduces $A$'s uncertainty regarding the distribution of $LR_B$ under $H_1$ and $H_2$.

Suppose the expert, in addition to providing $LR_B$ for the case at hand, also provides results from validation testing where the expert was asked to evaluate $LR$s in reference scenarios where a third party knew whether or not the glass samples being evaluated were from the same source. Suppose there are $n_1$ $LR$s evaluated in scenarios that $A$ views as having come from the same distribution as $LR_B$ would have if $H_1$ were true, and that the logarithms of these $LR$s have a sample mean of $\bar{y}_1$ and a sample variance of $s_1^2$. Taking these validation test results into consideration reduces $A$'s uncertainty regarding the distribution of log $LR_B$ values under $H_1$. In particular, an application of Bayes' rule yields that $A$'s

---

[14] The exhibited behavior where $LR_A$ shrinks towards 1 for extreme values of $LR_B$ is a consequence of the priors chosen for the distribution of $LR_B$ under $H_1$ and $H_2$. Several different combinations of parameters under the normal-gamma distribution exhibited this behavior. Though somewhat surprising, this effect is irrelevant to the point of this example, which was chosen for its computational simplicity.

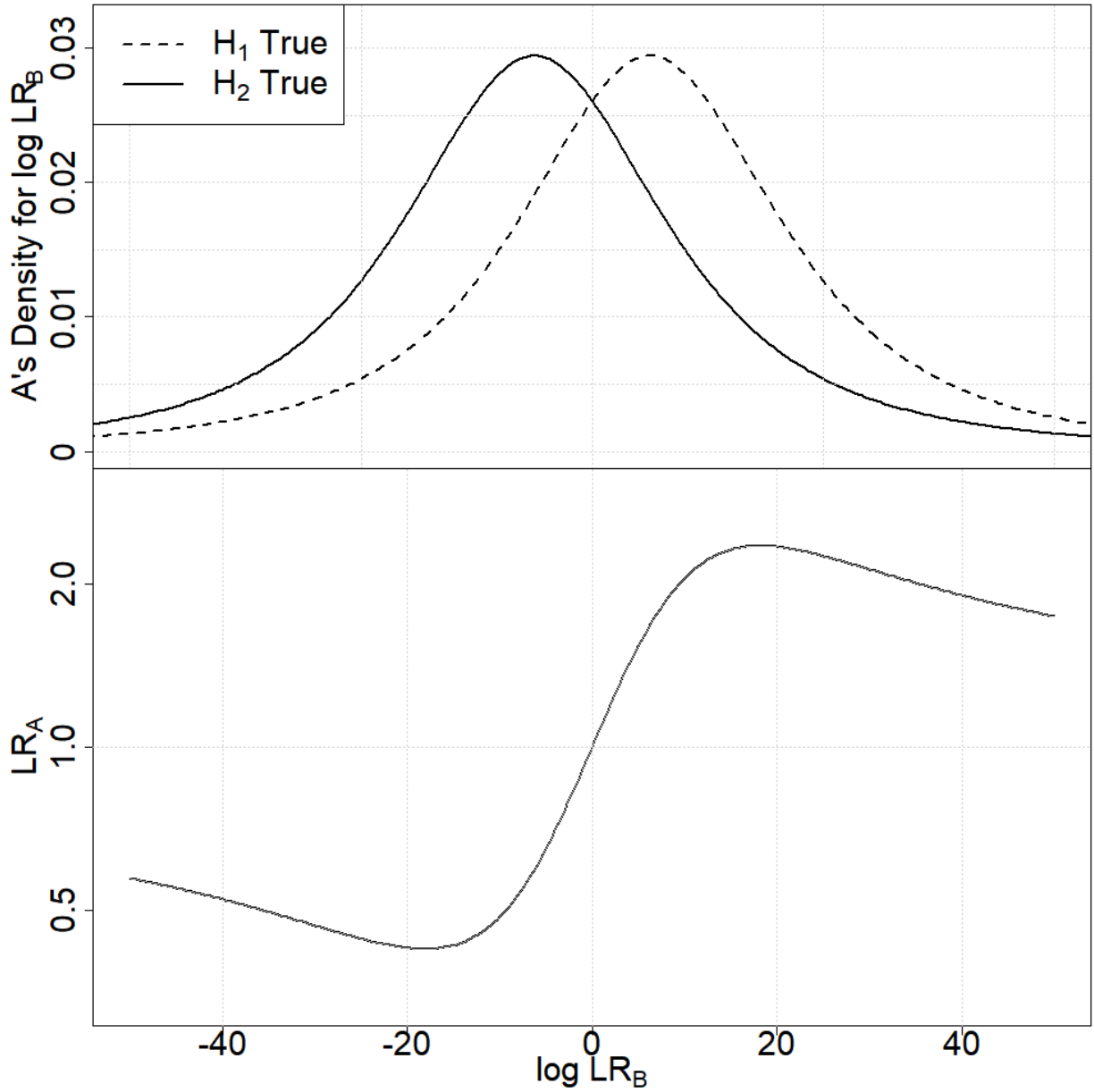

Figure 2: Top: Prior distributions assigned by $A$ for $LR_B$ under $H_1$ (dashed line) and under $H_2$ (solid line). Bottom: Resulting $LR_A$ values as a function of $log(LR_B)$.

updated uncertainty would follow a normal-gamma distribution with hyperparameters:

$$\mu_1^* = \frac{n_{\mu 1}\mu_1 + n_1 \bar{y}_1}{n_{\mu 1} + n_1}, \tag{2}$$

$$n_{\mu 1}^* = n_{\mu 1} + n_1 \tag{3}$$

$$n_{\tau 1}^* = n_{\tau 1} + n_1, \text{ and} \tag{4}$$

$$\frac{n_{\tau 1}^*}{\tau_1^*} = \frac{n_{\tau 1}}{\tau_1} + n_1 s_1^2 + \frac{n_{\mu 1} n_1 (\bar{y}_1 - \mu_1)^2}{n_{\mu 1} + n_1} \tag{5}$$

The effect of this additional information on $A$'s uncertainty regarding the distribution of $\log LR_B$ under $H_1$ is shown in the top panel of Figure 3 (dashed curves) for various numbers of validation tests, with each considered sample having a mean of 8 and a variance of 25. As one would expect, the larger the collection of validation samples provided to $A$ the more strongly $A$'s distribution for $\log LR_B$ under $H_1$ is pulled toward a normal distribution with mean 8 and variance 25.

The solid curves in the top panel of Figure 3 reflect a parallel exercise that considers the effect of providing $A$ with $LR$s from validation tests conducted in scenarios that $A$ views as having come from the same distribution as $LR_B$ would have if $H_2$ were true. For these computations the validation samples were considered as having a mean of $-12.5$ and a variance of 25.

The bottom panel of Figure 3 depicts the value of $LR_A$ as a function of $\log LR_B$ following disclosure of various numbers of validation $LR$s. As expected, providing additional validation test results increases the potential effect that $LR_B$ can have on $A$'s uncertainty regarding the truth of $H_1$.

As mentioned above, a Bayesian recipient does not require that the expert provide a likelihood ratio in order to apply Bayes' Rule. Suppose that $B$ provides a categorical conclusion such as "identified (type-I association)" [15]. In this instance, $A$ will need to evaluate how likely the expert is to report "identified (type I association)" under $H_1$ and under $H_2$, respectively. Again assuming that $A$ has no previous exposure to validation data regarding the performance of the examiner, $A$ may have substantial uncertainty regarding how frequently various conclusions are offered under various scenarios. For notational simplicity, let $p$ and $q$ denote the probability that the examiner will conclude "identified (type-I association)" under $H_1$ and $H_2$, respectively. Suppose in this instance, $A$ represents his/her uncertainty regarding $p$ and $q$ using a uniform distribution over the portion of the unit square where $p > q$. Without any additional information to further refine $A$'s uncertainty regarding the distribution of expert's conclusions under $H_1$ and $H_2$, $A$ would compute the likelihoods of hearing "identified (type-I association)" as 2/3 under $H_1$ and 1/3 under $H_2$, leading to a likelihood ratio of 2.

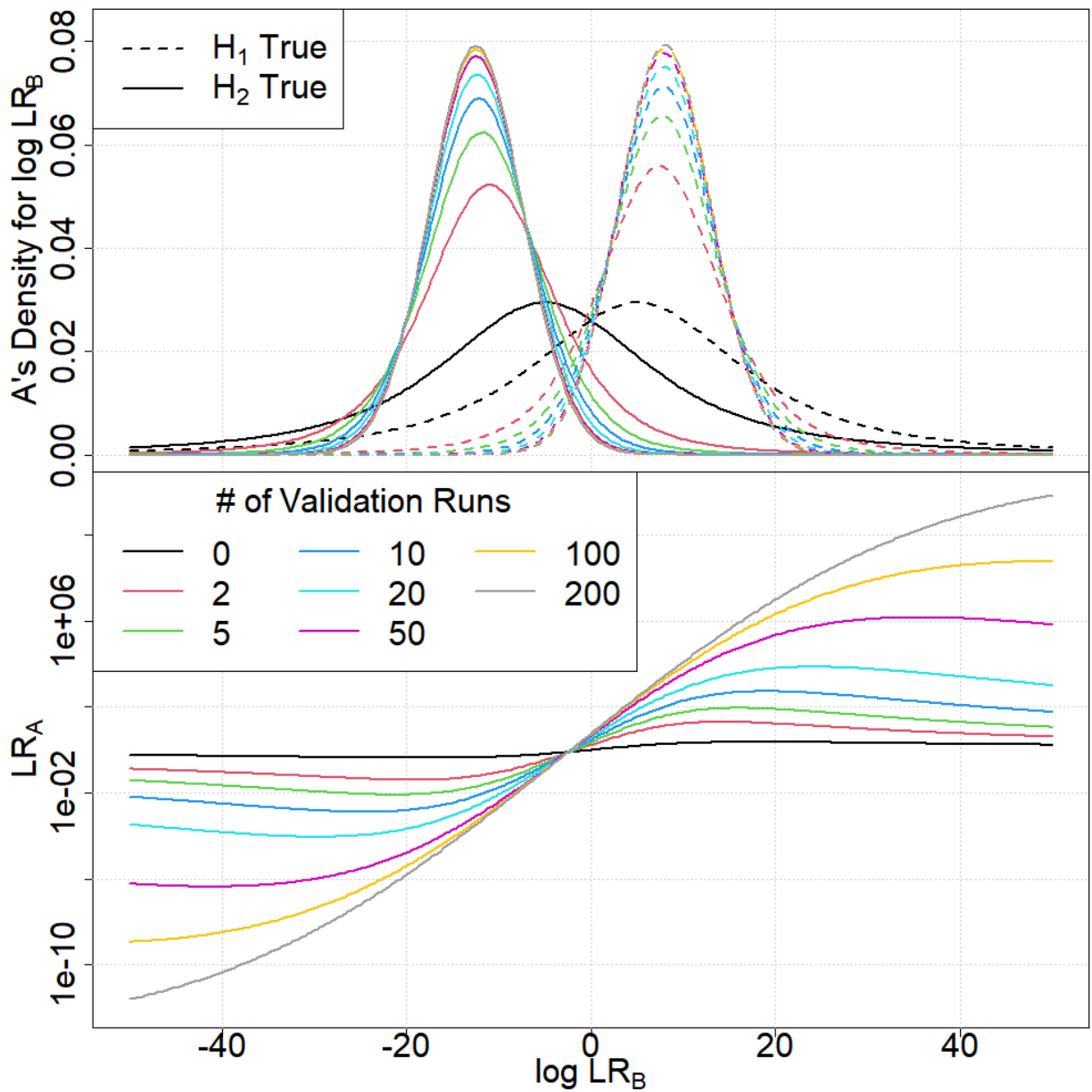

Figure 3: Top: Distributions for $LR_B$ under $H_1$ (dashed curves) and under $H_2$ (solid curves) for varying numbers of validation tests. Bottom: Value of $LR_A$ as a function of $log(LR_B)$ after disclosure of results from varying numbers of validation tests.

For the expert's conclusion to have a stronger influence on $A$, s/he must provide additional information that will lessen $A$'s uncertainty regarding the distribution of conclusions under $H_1$ and $H_2$. Suppose the expert provides validation testing results from $n_1$ and $n_2$ tests under scenarios $A$ views as representative of $H_1$ and $H_2$, respectively. Suppose further that "identified (type-I association)" was the provided conclusion in 95% of the $H_1$ true scenarios and 5% of the $H_2$ true scenarios. Figure 4 depicts the effect this additional information has on $A$'s assessment of the $LR$ corresponding to the expert's provided conclusion for various numbers of validation tests. As expected, the more validation test results are provided, the closer $LR_A$ gets to 19 (=0.95/0.05).

The Bayesian recipients in the previous examples exhibit an initial degree of skepticism that limits the influence of an expert's opinion. The skepticism fades away as additional empirical results are offered to support the interpretation. This model of communication with a recipient continuously rewards additional empirical studies into the methods used by forensic experts, and there is no threshold of data beyond which Bayesian recipients accept the opinion of the expert as if it were their own. The Bayesian does not ask whether enough testing has been conducted, but assesses his/her personal uncertainty after considering how much testing has been done. This perspective affects how the validation process or black box studies are viewed. In particular, the notion of a method being "validated" does not make sense. Each bit of testing provides a little more insight into the performance of the studied method, but at no point is the knowledge complete. There is always an ongoing cost-benefit trade-off for additional testing. We believe the value of validation and black box studies is not to justify that expert opinions be offered and taken literally, but to provide the option for recipients to rely on empirical results to inform their own interpretation of the offered opinion or result.

---
[15]see http://www.forensicsciencesimplified.org/trace/TraceEvidence.pdf

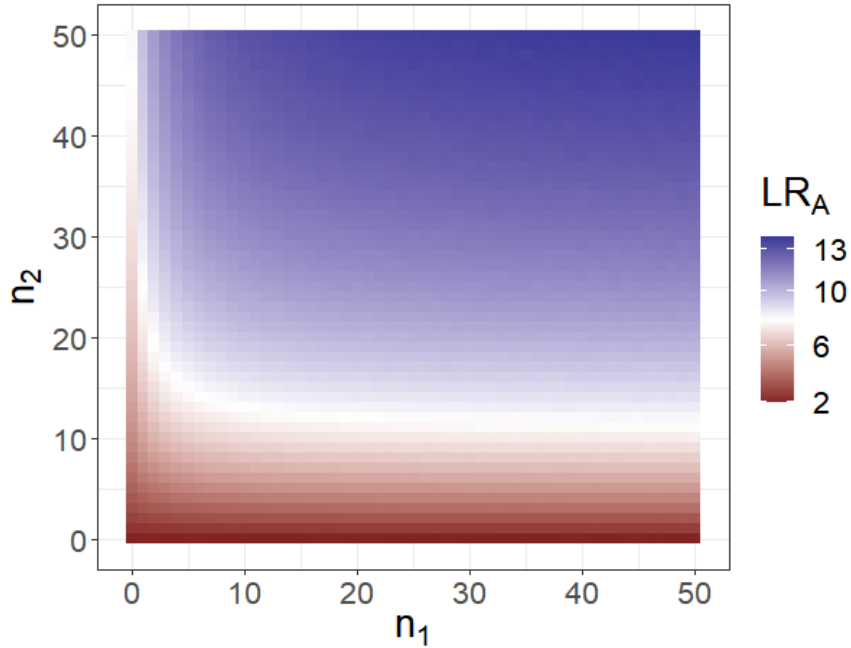

Figure 4: Heatmap showing the effect of examiner conclusion on $LR_A$ after $A$ is provided with results of various numbers of tests performed under $H_1$ true scenarios ($n_1$ tests) and $H_2$ true scenarios ($n_2$ tests).

In reality, most validation tests or reference cases will deviate from the current case in some less than ideal form. The extent to which one feels performance on reference compar-isons is indicative of performance for the current case is personal, which is a reason the data itself, rather than someone's interpretation of that data, must be directly presented and discussed. Additionally, not all cases within a given forensic discipline are equally challenging. For instance, some latent prints are clearer than others. The distribution of outputs can also be affected by factors other than which of the propositions of interest is true. As with error rates, it is critical to establish a collection of factors believed to affect the performance of a given method and consider the relevance of available reference comparisons to the case at hand in light of these factors.

Many recipients will not be familiar with formal Bayesian reasoning and some may feel that the implications from this section do not apply to these recipients. Of course, recipients do not need to explicitly follow Bayesian formulas for the punchline to hold.

## 3.2 Practical Implications

Even though it is unlikely that a recipient would actually conduct an analysis like those in Section 3.1, applying Bayesian reasoning to evidence communication still leads to relevant takeaways. In particular, we believe a legal system that aspires to have triers of fact and other stakeholders make rational and logical decisions should pursue communication practices intended to accomplish the following goals:

- Experts should avoid communicating in a manner that suggests recipients should accept an expert's opinion as their own. Recipients should not expect an expert to tell them what to think. Recipients should also not be made to believe that, due to the complexity of the math and the underlying science, they would be unable to assess a value of presented evidence for themselves.

- Experts and recipients alike should seek the meaning of a particular offered opinion

not by the literal meaning of the opinion, but by looking at performance data for the method used to form the opinion. As mentioned earlier, we expect most non-specialists are capable of understanding performance data and could use that knowledge to make judgments on their own.

- Experts and recipients should be mindful of the range of opinions. Recipients should especially keep this in mind when considering what a single expert's opinion means to them. However, this becomes impossible when no information is given to them regarding the range of plausible opinions.

While we arrive at these points by considering Bayesian reasoning in the context of evidence communication, similar points have also been recognized in the legal community. The first bullet point is closely related to concerns with a jury's deferential behavior towards experts as repeatedly mentioned in the U.K. Law Commission Report (https://assets.publishing.service.gov.uk/government/uploads/system/uploads/attachment_data/file/229043/0829.pdf). The U.K. Law Commission Report also emphasizes the importance of acknowledging a range of opinions. The NAS Report (2009) and the PCAST Report (2016) both highlight the central role of empirical performance data in assessing method reliability. We view method reliability as the extent to which an output from that method may impact one's belief (analogous to the effect of performance data on the Bayesian recipient's uncertainty as discussed in the previous section). The Daubert checklist (https://www.law.cornell.edu/rules/fre/rule_702) also explicitly mentions the role of testing and error rate information in reliability assessments of expert testimony.

The second and third bullet points are potential means to help accomplish the first. One way to encourage recipients not to simply defer to the expert is for each expert who is providing an opinion to also provide results from validation testing to show what opinions s/he reported during ground truth known scenarios. The expert would then explain that this data provides a basis for recipients to form their own assessments of the opinion offered in the current case. This was discussed in the Bayesian recipient section above.

Another way to dissuade deferential behavior is to discuss the existence of a range of reasonable opinions for interpreting any data and that any specific opinion is just one value from this set. While a complete range of reasonable opinions is generally unknowable in any given instance, discussing multiple opinions draws attention to the range rather than any one specific opinion. From our perspective, emphasizing a specific opinion, without instructing recipients to find their own meaning of that opinion by looking at performance data, carries the risk of recipients misunderstanding the expressed opinion to be a logically and scientifically-supported recommendation for how they should feel. This may lead to a misconception that any reasonable interpretation of the available information would not differ substantially from the offered opinion. If the expert used statistical modelling as part of their interpretation, additional models could be applied and the results from those models could be used to begin to shed light on how fragile or robust the offered opinions are to modeling assumptions.

Discussing a range of opinions could help to clarify the limits of expertise. While any expert has a limited domain of expertise, it can be difficult for laypersons to identify those limits for a given expert. Recipients may assume that whatever is presented by the expert falls within the expert's domain of expertise. It is our view that there is always a range of reasonable opinions and, as discussed in the following section, the choice of one specific opinion from within that range is a matter of personal preference, not expertise.

# 4 A Statistician's Expertise and Limits

Our focus on the additional information that should accompany expert opinion originates from considering the limits of expertise for a statistician or any person advising others on probabilistic interpretation. Probabilistic interpretations may have increased risk of receiving undue credence because many recipients may be unaware that probabilities are also opinions. In addition, many recipients may be unfamiliar with, or even intimidated by, statistical computation and may feel overly deferential to the expert. While people may generally understand that different models give different answers, we feel there is a strong risk that recipients, unless explicitly informed otherwise, will presume that an expert's interpretation is the "best" or "most appropriate." We feel this is a consequence of failing to clarify the limits of a statistician's expertise that result from the fact that probabilities are personal. As indicated above, the notion that probabilities are personal has been covered by many authors (de Finetti, 2017; Lindley, 2013; Kadane, 2020), and in this section we reflect on what implications this has for expert communication.

The topic of domain of expertise has been addressed by many authors, especially in the context of experts providing categorical conclusions (e.g., Aitken, Roberts, *et al.*, 2010; Evett *et al.*, 1987). For example, consider the phrase "The bullet recovered from the victim's body was discharged from this firearm." Among forensic statisticians, there is general agreement that this opinion is inappropriate because it requires explicit consideration of prior probabilities and the implications of a given decision, both of which are considered to be outside an expert's domain of expertise. The issue is that providing conclusion opinions risks giving the false perception that all components of inference up to the conclusion are within the expert's domain of expertise, leading to increased risk that recipients will give undue weight to the offered opinion and may accept it at face value without sufficient basis. Many authors recommend that experts summarise their findings using likelihood ratios to avoid elements that are outside the expert's domain of expertise (i.e., prior probabilities, decisions, and their associated costs). This section presents a stricter view of the limits of expertise for a statistician. In particular, we argue that the act of choosing of a single model from a pool of justifiable models is also outside the domain of expertise for any expert, including statisticians.

Before additional data is considered, the collection of candidate models that could be used to represent the considered evidence is very large (in fact, infinite). As more and more data, representative of the mechanism that produced the evidence, becomes available, some of the candidate models will be judged to be implausible and will be removed from the pool of plausible models, causing the pool to shrink. At the end of data collection, when all available data has been considered, there is still a pool of models, each of which could plausibly describe how that data might have been generated. These are models that cannot be filtered out by any of the usual diagnostic tests statisticians apply.

Ideally, one would like to consider the implications on the question of interest (for instance, a strength of evidence assessment or a prediction of a future value as in forecasting) of each of the surviving models in the pool. One could then report what is known about the collection of values one could obtain for a forecast or a strength of evidence assessment based on each model in the pool. This would reflect full awareness of all plausible inferences for a given body of data. Unfortunately, no one, not even a statistician, can implement or even consider all possible models, so no one has full awareness of all plausible inferences. Thus, providing an absolute bound on the uncertainty of a probabilistic interpretation (other than the definitional limits such as probabilities must be between 0 and 1, and probability densities and likelihood ratios cannot be negative) is outside the statistician's domain of expertise.

While no one can consider all possible models, statisticians have the tools to recognize large collections of models that may all be plausible candidates. The more training a statistician has, the larger the pool of candidate models s/he can, and should, identify and implement. Similarly, statisticians are also familiar with tools to help judge if the data are inconsistent with any of the candidate models. Statistician can, and should, use these tools acquired from their training and experience to arrive at a collection of plausible interpretations. The more training a statistician has, the larger the pool of candidate models s/he can demonstrate to be implausible, as well. Ultimately, results based on the largest possible pool of plausible models can, and should, be provided to recipients to help them shape their understanding of the range of reasonable opinions.

Because a statistician will still only have explored a subset of all plausible models, however, the fact remains that other plausible models may produce values outside the range of values presented by the statistician. This would be a fair characterization of what the statistician can infer based on available facts and data and marks the limits of the statistician's domain of expertise. Consequently, an expert emphasizing a specific probabilistic interpretation (e.g., a particular likelihood ratio value), we argue, also oversteps the boundaries of their expertise.

If a statistician or other expert emphasizes a single value for a forecast or for a strength of evidence assessment, s/he has, in effect, chosen an arbitrary value from the pool of plausible values. This would not matter if it were somehow known that any value from the pool of values would lead to the same outcome in the case under review, but this is generally unknowable since no statistician can consider all models. The key observation is that, after arriving at a pool of plausible models or a pool of plausible values for strength of evidence, there is NO further expertise possessed by the expert. The choice of a single value from a pool of plausible values is not a matter of expertise; instead, perhaps, it could be argued to be an effort to simplify communication but it is a simplification that comes at the cost of misrepresenting capabilities the expert garners from math and science training. Candidly discussing the range of reasonable opinions provides the trusting recipient a way of assessing the reliability of the expert's opinion. The Royal Commission report titled "Expert Evidence in Criminal Proceedings in England and Wales" (https://assets.publishing.service.gov.uk/government/uploads/system/uploads/attachment_data/file/229043/0829.pdf) discusses this issue in great detail (paragraph 5.35 in particular). As previously described above, however, a good solution to this problem is to prompt recipients to view the expressed opinion as a 'score' by focusing on empirical validation results and form their own understanding of the offered opinion and its reliability.

# 5 Conclusion

By exploring how a Bayesian fact finder might process information received from one or more experts together with information revealed during cross examination of the expert, we gained valuable insights into the benefits of making available to fact finders, performance data regarding expert's process of formulating opinions. By considering a lay person recipient of expert information, and how s/he might misjudge the expert's limits of expertise, we were led to consider, in particular, what the domain of expertise is for experts trained in probability and statistics and to recognize when they step outside their domain.

The goal of experts is to provide information that would benefit fact finders and others in the judicial system who are tasked with making a decision. Many influential forensic scholars are of the view that this is accomplished by providing their own $LR$ to fact finders and other decision makers. Bayesian/logical reasoning is often stated as the basis for this view. Some have even attempted to give this view a 'normative' status. While it is generally accepted that a rational thinker MUST update their prior beliefs to obtain their posterior beliefs through the use of their own $LR$, there are, however, no such 'normative' arguments for promoting communication of evidence from the expert to a decision maker via an expert's likelihood ratio. The practice of using an expert's likelihood ratio as your own represents a departure from Bayesian reasoning. The specific personal opinion of one person, even an expert, is not a suitable stand-in for informing another individual who is expected to make rational decisions. Expert communication should strive to focus on relevant facts, with appropriate and voluntary disclosure of any injections of personal opinions by the expert. Interpretations or conclusion statements should be recognized by all stakeholders simply as the outcome of a comparison process rather than as the answer key for understanding the presented evidence. The anticipated potential impact of any report or testimony should be limited by the amount of provided data that illustrates the effectiveness of the chosen comparison process in cases like the one at hand.

**Disclaimers:**

**Acknowledgements:**

We thank Will Guthrie, Jan Hannig, Martin Herman, and Yooyoung Lee for their valuable comments on an initial draft of this paper.

# Appendix-A
# Probability is Personal

Probability theory is, indeed, extremely useful for an individual who has to make decisions in the presence of uncertainty. Unfortunately, one person's probability does not transfer to another person because how uncertain one person feels regarding the truth of a proposition is not necessarily how uncertain someone else should, or would, feel about it. The feeling of uncertainty is highly subject specific. Different individuals can (and generally do) have different degrees of belief about the same event or proposition, even if they have the same background data or information available. This is because data, by themselves, do not produce probabilities.

It has been recognized by many of the founders of modern probability theory that probabilities are personal (Lindley, 2013; Kadane, 2020; de Finetti, 2017). According to them, a probability is a quantitative expression of the degree of belief in the truth of a statement (proposition, hypothesis, event) that an individual has based on their knowledge and other beliefs. Kadane says, in the very first chapter of his book titled "Principles of Uncertainty" the following:

> "Before we begin, I emphasize that the answers you give to the questions I ask you about your uncertainty are yours alone, and need not be the same as what someone else would say, even someone with the same information as you have, and facing the same decisions."

The only requirement for a logical and mathematical treatment of such personal probabilities is that the collection of probabilities assigned by an individual to a set of related propositions obey the basic laws of probability. This property is often referred to as *coherence*.

The basic laws of probability themselves can be derived by adopting the "avoid sure loss"

principle (Kadane, 2020), that is, one will not make a bet that is known, with certainty, would result in a loss to the person making the bet. Using this argument (sometimes also referred to as a "Dutch book argument") the fundamental laws of probability can be derived. As Kadane says (words in italics added by us),

> Avoiding being a sure loser requires that your prices *(probabilities)* adhere to the following equations:
>
> (1.1) $Pr\{A\} \geq 0$ for all events $A$
>
> (1.2) $Pr\{S\} = 1$, where $S$ is the sure event
>
> (1.3) If $A$ and $B$ are disjoint events, then $Pr\{A \cup B\} = Pr\{A\} + Pr\{B\}$.
>
> If your prices [(i.e., probabilities)] satisfy these equations, then they are coherent.

Kadane goes on to say,

> Coherence is a minimal set of requirements on probabilistic opinions. The most extraordinary nonsense can be expressed coherently, such as that the moon is made of green cheese, or that the world will end tomorrow (or ended yesterday). All that coherence does is to ensure a certain kind of consistency among opinions. Thus an author using probabilities to express uncertainty must accept the burden of explaining to potential readers the considerations and reasons leading to the particular choices made. The extent to which the author's conclusions are heeded is likely to depend on the persuasiveness of these arguments, and on the robustness of the conclusions to departures from the assumptions made.

In particular, being coherent does not imply being true.

A compelling illustration of the fact that probabilities, and hence likelihood ratios, are personal, is offered in the book (Kadane and Schum, 2011) in which the authors use the case involving "... a shoemaker named Nicola Sacco and a fish peddler named Bartolomeo Vanzetti who were charged with first-degree murder in the slaying of a payroll guard during an episode of armed robbery that took place in South Braintree, Massachusetts, on April 15, 1920, to illustrate the multiplicity of issues that arise when considering a complex collection of evidential material and attempt to derive probabilistic conclusions using a chain of plausible arguments. ... " In particular they illustrate (see Chapter 6 of their book) to what extent likelihood ratio assessments made by the authors, and another individual very familiar with the details of the Sacco and Vanzetti case, differ from one another.

In reality, different people may assign different values to probabilities for a variety of reasons they consider to be valid. This is not an issue when each individual is making probability assessments for use in their own decision making, but it is an absolutely critical issue if attempting to tell someone else what their uncertainty should be. Most types of evidence are complex structures with many attributes, each of which could be considered to varying degrees or ignored altogether by different individuals. A concrete example is provided by the different approaches that are still being practised in DNA mixture interpretation: the binary model, the semi-continuous model, and the continuous model, to name a few, where some models use only a part of the information used by other models. That is, even $E$ (i.e., evidence) by itself is rather ambiguous.

Further, terms placed to the right of the vertical bar, such as $I$ in $Pr[E|I]$, represent information considered as indisputable fact by the individual forming the probability; however, individuals may disagree as to what constitutes fact. Just because one person, even an expert, treats something as a proven fact, does not mean all decision makers must. A decision maker might agree with some portions of what an expert treats as factual and question, or outright reject, other portions. It makes good sense to consider probabilistic models that accommodate these real world situations. The authors of `AN` are ignoring these, very real, considerations in embracing the practice of having decision maker $A$ use $LR_B$ in place of their own $LR$ (i.e., $LR_A$).

For the benefit of readers who have not been previously exposed to the fact that probabilities are personal, we provide an illustrative example below.

**Example.** Suppose a coin was tossed by a mechanical device eight times and the results were $HHHHHTTT$, in this order. Let us now consider the question "What is the probability that the result of the ninth toss would be heads ($H$)?" We illustrate that answers to this deceptively simple question are personal by considering the responses of three different hypothetical individuals, say A, B, and C.

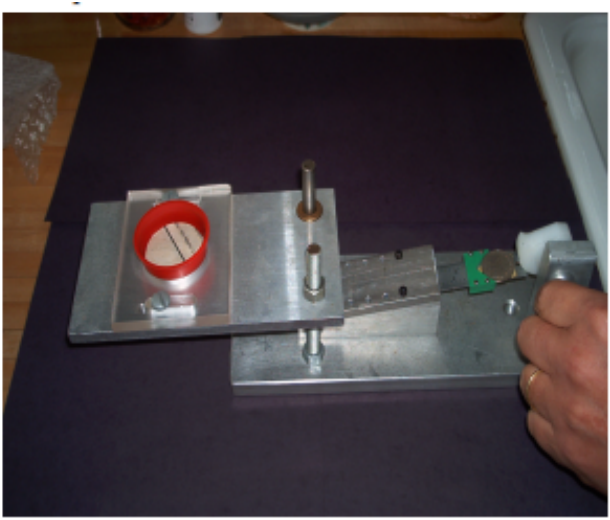 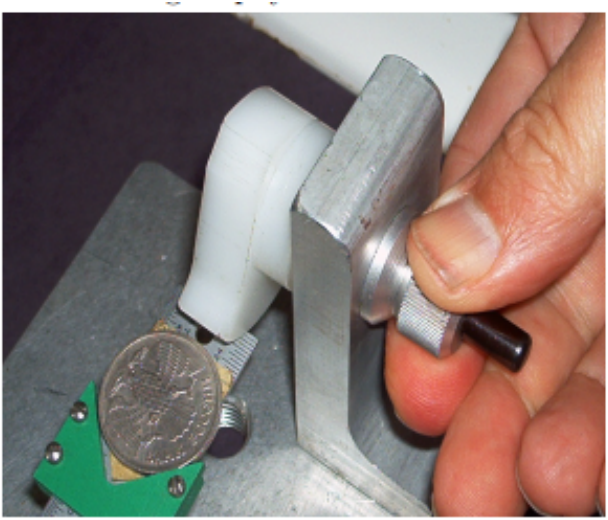
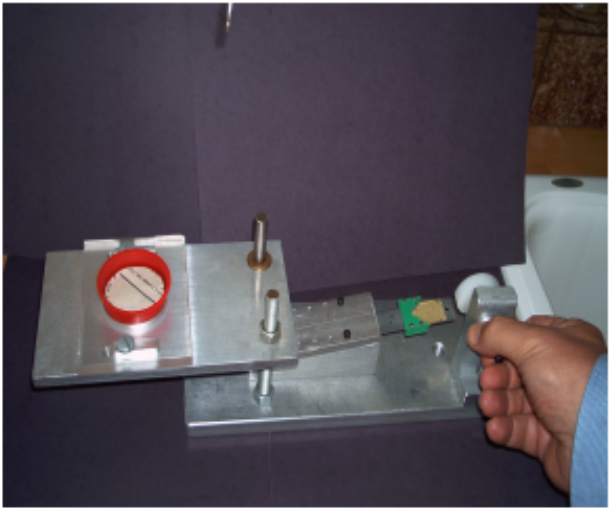 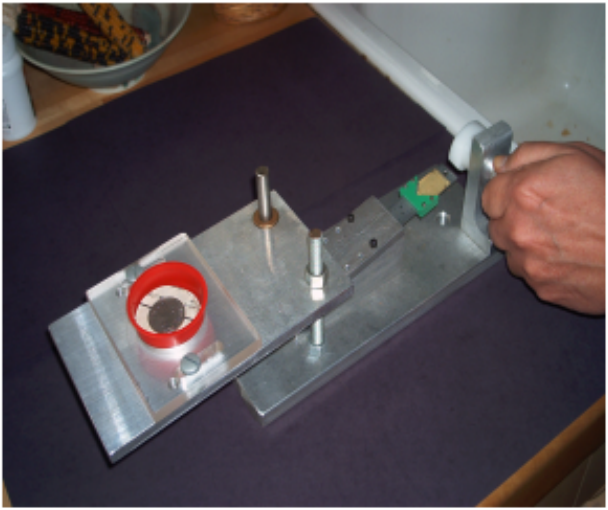

Figure 5: Mechanical coin tossing device used by J. B. Keller. [Keller, 1986] The probability of heads, American Mathematical Monthly, 93:191-197.

Individual $A$ believes from the outset that the coin tossing mechanism will be fair and the tosses will be independent and assesses the probability of the ninth toss being heads to be 1/2, regardless of what was observed among the first eight tosses. She may even carry out a diagnostic statistical test to assess whether the observations are inconsistent with her assumed model and note that there is no compelling evidence against her model.

Individual $B$ feels uncertain about the behavior of the coin-flipping apparatus and rep-resents his uncertainty using a uniform distribution for $p = Pr(Heads)$. He further assumes that, if $p$ were known, then the outcomes of individual tosses follow an independent and identically distributed Bernoulli model with probability $p$ of obtaining 'heads' in each toss. In particular, the number of heads observed in $n$ flips would follow a binomial distribution with parameters $n$ and $p$. After observing that 5 out of 8 flips resulted in heads, B's uncertainty regarding $p$ follows a beta distribution with parameters $\alpha = 6$ and $\beta = 4$. The expected value of this distribution is 0.6, and $B$ assigns a probability of 0.6 to the event that the ninth toss will be heads.

Individual $C$ views the tossing device and wonders if it might be prone to "drifting" such that the forces applied for tossing the coin keep changing gradually from one toss to the next. To account for this possibility, $C$ does not assume the tosses will all be independent, but instead chooses to represent the probability of heads on a given toss as being dependent on the outcome of the previous toss. More specifically, $C$ conceptualizes the flipping system using two separate probabilities, $Pr$(Next flip heads|previous flip was heads) $= p$ and $Pr$(Next flip heads|previous flip was tails) $= q$. Furthermore, $C$ represents her uncertainty about $p$ and $q$ using (mutually independent) uniform distributions. Among the last seven flips of the observed sequence $HHHHHTTT$ there are clearly four heads and one tails among the five flips that immediately follow an observed heads, and two flips that immediately follow an observed tails, both of which are tails. There is some ambiguity regarding what to do with the first flip, which was heads, because $C$ does not know the outcome of the flip that occurred before it, which we denote $Y_0$. $C$ reflects this uncertainty by assuming $Y_0$ was as likely to have been heads as it was tails. After applying Bayes' rule, the updated uncertainty regarding $p$ is the average of two beta distributions, one with parameters $\alpha = 6$ and $\beta = 2$ (reflecting the instance where $Y_0$ was heads and the first heads in the observed sequence is included in the total) and the other with parameters $\alpha = 5$ and $\beta = 2$ (reflecting the instance where $Y_0$ was tails and the first heads in the observed sequence is not included in the total). Similarly, the updated uncertainty regarding $q$ is the average of two beta distributions, one with parameters $\alpha = 1$ and $\beta = 3$ (reflecting the instance where $Y_0$ was heads and the first heads in the observed sequence is not included in the total) and the other with parameters $\alpha = 2$ and $\beta = 3$ (reflecting the instance where $Y_0$ was tails and the first heads in the observed sequence is included in the total). Because the last observed flip in the observed sequence of flips was tails, C's probability that the ninth toss will be heads is given by the expected value of $q$, which is 0.325.

Each of the three individuals above has applied Bayes' rule correctly and therefore each of the three individuals can claim to be logical and coherent. Yet the perceived probability of heads in a ninth flip differs substantially across the three individuals and none of them can be labeled as incorrect. Even though all individuals have the same knowledge of the empirical data (the results of the first eight flips), they arrive at different personal probabilities because their initial beliefs were different. None of their respective mental stories regarding how the mechanical device might behave is inherently more truthful than any other. Correspondingly, none of their chosen priors are more appropriate than any other.

This example is intended to illustrate the basic fact that, even in simple scenarios, different individuals can follow Bayesian reasoning and arrive at different probabilities for the same propositions given the same data. The more complex a statistical model becomes, the more opportunities there are for modeling choices to substantially affect the outcomes of model.

In general, the subjectivity of probabilistic interpretation has many sources (e.g., confidence in the motives and skills of the persons collecting and processing the evidence in this case, the representativeness of reference evaluations used to inform distributional assumptions for the given case, the actual assumed distributions, etc). This subjectivity of probabilistic interpretation should influence our choice in evidence communication strategies. If we know that interpretations vary across individuals and models, why would we choose to emphasize the interpretation of one expert or one model, especially without thoroughly attempting to understand the level of variability among experts or models in a given case? In the example above, we suggest that the informational value is entirely contained in the data $HHHHHTTT$ and any background information that might be available regarding the coin tossing device. Hearing the personal interpretation from one person (e.g., who thinks the probability that the next flip results in heads is 0.5 (or 0.6, or 0.325)) does not add any scientifically defensible value. In fact, focusing on a single probabilistic interpretation can be misleading since it does not convey to the recipient that there are many other plausible and equally justifiable assessments. The recipient is left with inadequate information to judge the reliability of the given opinion and, in many cases, they may not even be aware of this fact.

# Appendix-B
# Addressing Responses to Lund and Iyer (2017)

We first expressed concerns regarding the hybrid modification of Bayes' rule in our 2017 paper, titled "Likelihood Ratios as Weight of Forensic Evidence: A Closer Look," henceforth referred to as `LI`. Among the published rebuttals, Gittelson *et al.* (2018) agreed that the hybrid modification of Bayes' equation is invalid. In particular, the authors explicitly state "To update their prior odds to their posterior odds, a DM must assign their own LR." We find comfort in the apparent agreement, at least in principle, on this important point. However, they also characterize the concerns we raised regarding the hybrid Bayes' equation as a straw man argument, writing "No one advocating the Bayesian approach to evidence evaluation has ever argued against the self-evident truth that DMs update their prior odds with their own personal *LR*s to obtain their own personal posterior odds." The paper also cites a response from Geoffrey Morrison (Morrison, 2017) that also claims our concern regarding the hybrid modification of Bayes' rule is a straw man argument.

While the primary purpose of the main text was to illustrate how a recipient could apply genuine Bayesian reasoning (as opposed to the hybrid modification) in response to learning an expert's likelihood ratio (or other forms of opinions), we feel it is also important to reiterate why we disagree with the characterization that our concerns about the hybrid modification to Bayes' rule in the context of forensic evidence interpretation are based on a straw man argument. As stated in the introduction, we believe the issue is more subtle than experts or academics explicitly saying people should not update their personal prior odds with their own personal $LR$s. It is our impression that many proponents of experts communicating by providing LRs overlook a recipient's personal $LR$ and implicitly supports the hybrid modification to Bayes' rule by omission. We provided several illustrations of this type in the Introduction, the last two of which came from an article whose author list includes eight of the authors of Gittelson *et al.* (2018). We repeat those excerpts here for the reader's convenience:

- "Bayes' Theorem provides a model that clearly distinguishes the role of the scientist and that of the fact finders. *The role of the scientist is to advise the fact finders on the strength of the evidence by assigning the LR. Any consideration of the prior or posterior odds (or the probability) of the propositions is left to the fact finders.*" (Buckleton *et al.*, 2020)

- "*The role of the forensic scientist is to assign the probabilities of the evidence given the propositions that are considered.*" (Buckleton *et al.*, 2020)

These phrasings from 2020 appear to be a far cry from the unequivocal repudiation of the hybrid modification to Bayes' rule announced in the rebuttal by Gittelson *et al.* (2018). We hope their future discussions of likelihood ratios and Bayes' rule, in publications, reports, and testimony, will return to the clarity on this point exhibited in Gittelson *et al.* (2018).

We further provided examples of more explicit support for the hybrid modification to Bayes' rule. Authors from papers containing two of the quotes (provided below for the reader's convenience) were also involved in writing two responses to `LI` that attempt to support the use the expert's $LR$ in a hybrid application of Bayes' equation.

- "The main focus of attention will be confined to the perspective of how one can assess the value of scientific findings in order to inform about how findings should affect the views of others on selected issues in a case." (p. 182) (Biedermann, Taroni, *et al.*, 2014)

- "Determination of the BF (Bayes' factor) is typically considered to be in the domain of the forensic scientist." (Taroni *et al.*, 2016)

Aitken and Nordgaard (Aitken and Nordgaard, 2018) describe the practice of recipients accepting an expert's provided $LR$ value as their own, when applying Bayes' rule to obtain posterior probabilities, as "an excellent method in which information may be transferred from an expert to a decision maker." Hereafter, we refer to this article as `AN`.

In a second article (Aitken, Nordgaard, *et al.*, 2018 - hereafter referred to as `ANTB`), the authors responded to the statement in `LI` which said "we hope the forensic science community comes to view the $LR$ as one possible, not normative or necessarily optimum, tool for communicating to DMs (decision makers)." In their response, they revisit arguments of Good (Good, 1989a; Good, 1989b; Good, 1991) and (Aitken and Taroni, 2004) and then state "with some very reasonable assumptions, the assessment of uncertainty inherent in the evaluation of evidence leads inevitably to the likelihood ratio as the only way in which this can be done." `ANTB` fail to demonstrate that likelihood ratios are inevitable for experts communicating to decision makers about evidence. Rather, they simply shows that likelihood ratios are inevitable elements of logically updating *personal probabilities*, which was not disputed in `LI`. While we find the arguments in `AN` and `ANTB` unpersuasive, the viewpoints conveyed by these authors nevertheless clearly reflect their continued support for the hybrid modification of Bayes' rule.

A recent publication (Biedermann and Kotsoglou, 2020) cites the response papers (Gittelson *et* al., `AN`, `ANTB`) and describes them as having refuted key claims in `LI`. This is a mischaracterization of the situation. While we are aware that some influential forensic experts either do not agree with our concerns, or do not take those concerns seriously, by no means have our concerns been shown to be mathematically incorrect. On the contrary, the collection of responses, especially `AN` and `ANTB`, have only served to strengthen our earlier concerns. In the remained of this appendix, we examine the arguments of Aitken and Nordgaard (2018) and Aitken, Nordgaard, *et al.* (2018), both of which support the use of the hybrid modification of Bayes' rule, and conclude that their arguments have no relevance with regard to the concerns expressed in `LI`.

## Aitken and Nordgaard 2018

The first paragraph of `AN` states

> "We argue that although forensic scientists and fact finders have different information, no difficulty is caused in the evaluation process. This argument is particularly apposite given an argument in a recent article by Lund and Iyer (1). They argue that '[b]ecause the likelihood ratio is subjective and personal, we find that the proposed framework in which a forensic expert provides a likelihood ratio for others to use in Bayes' equation is unsupported by Bayesian decision theory, which applies only to personal decision making and not to the transfer of information from an expert to a separate decision maker, such as a juror'. It is argued here that this finding is not valid. The likelihood ratio provided by the forensic scientist is a valid method for the transfer of information from an expert to a separate decision maker."

The basic premise of `AN` is that, under certain conditions, a recipient accepting the likelihood ratio provided by the expert as their own will be led to the same posterior as would a single person who inherently has the combined information of both the recipient and the expert. What `AN` fails to articulate is that a likelihood ratio, like all probabilistic expressions, reflects a mixture of someone's account of information that might be considered factual by others as well as modelling choices that are unambiguously personal or subjective. This means that accepting someone else's likelihood ratio value as your own requires that you treat all the subjective choices that someone else made when forming the offered value as if they are facts. This is a departure from Bayesian reasoning, which requires that all parties interpret new information according to their own personal uncertainties. The hybrid modification to Bayes' equation supported by `AN` represents the recipient as fully certain that all their personal uncertainties will exactly correspond to the modeling assumptions made by the expert, before even having heard what those assumptions are. This misrepresentation of Bayesian reasoning is devoid of the necessary and natural step in which recipients critically reflect upon what the information provided by the expert means to them in accordance with their own uncertainties, as they existed before having ever heard from the expert.

If one adopts Bayesian reasoning and strives to make coherent assessments of one's own probabilities, then newly encountered information must always be processed with respect to one's own uncertainties in accordance with Bayes' rule. There is no exception for the scenario where the newly encountered information happens to be someone else's likelihood ratio, even an expert's. In terms of the communication model presented earlier, the expert interacts with evidence $E$ while considering propositions $H_1$ and $H_2$, forms perceptions $\phi_B(E_B, I_B)$, and articulates this in the form of $\mathcal{C}_\mathcal{B}(E_B, I_B)$ to $A$. This communication presumably includes the expert's $LR$ which is $LR_B = \dfrac{Pr_B[\phi_B(E_B, I_B)|H_1; I_B]}{Pr_B[\phi_B(E_B, I_B)|H_2; I_B]}$, along with any supporting information.

The recipient understands the expert's presentation as $E_A = \phi_A(\mathcal{C}_\mathcal{B}(E_B, I_B), I_A)$, which is the new information registered by the recipient. In more general scenarios, $E_A$ would also include information presented by other experts, also commenting on $E$, or information that arises during direct and/or cross-examination. Coherence requires that the recipient's probability assessments for $Pr_A[H_1|I_A; E_A]$, $Pr_A[H_2|I_A; E_A]$, $Pr_A[E_A|H_1; I_A]$, $Pr_A[E_A|H_2; I_A]$, $Pr_A[H_1|I_A]$, $Pr_A[H_2|I_A]$ must obey Bayes' rule

$$\frac{Pr_A[H_1|E_A; I_A]}{Pr_A[H_2|E_A; I_A]} = \frac{Pr_A[E_A|H_1; I_A]}{Pr_A[E_A|H_2; I_A]} \times \frac{Pr_A[H_1|I_A]}{Pr_A[H_2|I_A]}. \tag{6}$$

The factor by which $A$'s initial beliefs changed to their final beliefs is precisely their likelihood ratio $LR_A$ given by

$$LR_A = \frac{Pr_A[E_A|H_1; I_A]}{Pr_A[E_A|H_2; I_A]}, \tag{7}$$

which has been described as the 'value of evidence' (for $A$) and $log(LR_A)$ is the weight given by $A$ to the evidence as they understand it.

Throughout their article AN use notation like $Pr(E|H_1, I_A)$[16], which is generally read as "the probability of $E$ given the truth of proposition $H_1$ and background information $I_A$. Unfortunately, this representation can make it difficult to identify where the subjective elements of probability assignment are. The absence of a subscript for $Pr$ implies the function's arguments are sufficient to specify its output, regardless of who is doing the assessment. This can only be true if the subjective components from the assessor are tucked away in what they refer to as "background information," a phrase which does little to warn readers of the influence and consequences of the inescapable, and potentially impactful, subjectivity that accompanies probability assignment.

We now examine the conditions under which the practice of accepting the $LR$ value provided by an expert, as promoted in `AN`, reflects a valid application of Bayesian reasoning. This can only occur when $LR_A$ and $LR_B$ are equal. However, examining the expressions makes such instances seem exceedingly coincidental at best. To simplify discussion, consider, for instance, just the numerators of the two $LR$s. The numerator of $LR_A$ is $Pr_A(\phi_A(\mathcal{C}_B(E_B, I_B))|H_1, I_A)$ whereas the numerator of $LR_B$ is $Pr_B(\phi_B(E_B, I_B)|H_1, I_B)$. Two elements immediately jump out from comparing these two expressions:

- All components of $I_B$, including the expert's personal and subjective representations of uncertainty, have shifted from the left side of the vertical bar (as seen in $LR_A$) to the right side of the vertical bar (as in $LR_B$). That is, $A$ must extract all of $I_B$ from $\mathcal{C}_B(E_B, I_B)$, understand it, and unquestioningly accept it as true in its entirety, including all the modelling assumptions $B$ made to facilitate the calculation of their probabilities.

- In realistic scenarios involving multiple experts evaluating the same evidence or information revealed during cross examinations, $E_A$ can be expected to contain information that was not available to $B$ during their assessment of $E$.

---

[16] AN actually uses notation such as $Pr(E|H_1, I_a)$ and $Pr(E|H_1, I_b)$, but in this article we, instead, use $Pr(E|H_1, I_A)$ and $Pr(E|H_1, I_B)$ to align with the notation $A$ and $B$ used by us (and them) for the recipient and the expert.

If $LR_A$ and $LR_B$ are not equal, which would almost always be expected to be the case, then $A$ using $LR_B$ in place of $LR_A$ is a clear departure from Bayesian reasoning. This is why we reiterate our statement that "the proposed framework in which a forensic expert provides a likelihood ratio for others to use in Bayes' equation is unsupported by Bayesian Reasoning."

Although `AN` is advocating for a practice that departs from Bayesian reasoning, suppose, for the moment, that one is willing to accept its deficient framework. A careful scrutiny of the mathematical derivations within their proposed setting reveals erroneous arguments. In particular, we explain why the assumed setup in `AN` is unrealistic and why their subsequent arguments are mathematically incorrect.

Consider the setup described in `AN`.[17]

> Denote two participants in the judicial process as $A$ (fact finder) and $B$ (expert) and denote the background information available to each as $I_a$ and $I_b$, respectively, with the overall background information $I = I_a \cup I_b$. (8)

We now focus on specific statements made in `AN` and carefully examine their validity. We give the page number of the journal issue where `AN` appeared along with column and line numbers so that the reader will find it easy to locate the relevant sentences in `AN`.

---

[17]Note that `AN` uses lower case 'a's and 'b's in symbols such as $I_a$ and $I_b$ but we have been using upper case $A$s and $B$s such as $I_A$ and $I_B$ in this paper. In this appendix, however, we will use $I_a$, $I_b$, etc., for the sake of being consistent with the notation in `AN`.

**Page 648, last 4 lines of Column 1 and first two lines of Column 2**

> Partition $I = I_a \cup I_b$ into $I_{a \setminus b}$, $I_{b \setminus a}$ and $I_{ab}$ where $I_{a \setminus b}$ is background information that $A$ has but not $B$, $I_{b \setminus a}$ is background information that $B$ has but not $A$, and $I_{ab}$ is the background information common to $A$ and $B$. Thus, the elements of the set $\{I_{a \setminus b}, I_{b \setminus a}, I_{ab}\}$ are mutually exclusive and their union is $I$; $I$ may be written as $\{I_{a \setminus b}, I_{b \setminus a}, I_{ab}\}$ or $\{I_{a \setminus b} \cup I_{b \setminus a} \cup I_{ab}\}$. (9)

AN describes the symbol $I_b$ as the background information that $B$ has. In reality, $I_b$ represents information that $B$ *completely* believes to be true. In other words, $I_b$ is not necessarily factual information that would be accepted by all others. A similar comment applies to $I_a$. Then $I_{ab}$ represents information that both $A$ and $B$ *believe* to be true. This distinction between factual information versus statements that an individual believes to be true (but others may or may not have the same beliefs) is very important. For instance, when $B$ makes modeling assumptions to facilitate his/her probability calculations, such assumptions do not constitute information that all others will accept as factual. However, inferences made by $B$ are conditional on the assumptions made by $B$ (which need not be accepted by $A$ as being true) and hence probability assessments made by $B$ are influenced not only by factual information but also by beliefs that $B$ holds to be true that others may not hold to be true.

**Page 648, Equation (2)**

We will be referring to Equation (2) on page 648 of AN. This equation is shown in the box below for the convenience of the reader.

$$\frac{Pr[H_p|E,I]}{Pr[H_d|E,I]} = \frac{Pr[E|H_p,I]}{Pr[E|H_d,I]} \times \frac{Pr[H_p|I]}{Pr[H_d|I]}$$

$$\implies \frac{Pr[H_p|E,I_{a\backslash b},I_{b\backslash a},I_{ab}]}{Pr[H_d|E,I_{a\backslash b},I_{b\backslash a},I_{ab}]} = \frac{Pr[E|H_p,I_{a\backslash b},I_{b\backslash a},I_{ab}]}{Pr[E|H_d,I_{a\backslash b},I_{b\backslash a},I_{ab}]} \times \frac{Pr[H_p|I_{a\backslash b},I_{b\backslash a},I_{ab}]}{Pr[H_d|I_{a\backslash b},I_{b\backslash a},I_{ab}]}$$

$$\implies \frac{Pr[H_p|E,I_a,I_b]}{Pr[H_d|E,I_a,I_b]} = \frac{Pr[E|H_p,I_b]}{Pr[E|H_d,I_b]} \times \frac{Pr[H_p|I_a]}{Pr[H_d|I_a]}$$

$$\implies \frac{Pr[H_p|E,I]}{Pr[H_d|E,I]} = \frac{Pr[E|H_p,I_b]}{Pr[E|H_d,I_b]} \times \frac{Pr[H_p|I_a]}{Pr[H_d|I_a]} \quad (2) \quad (10)$$

The paragraph immediately following their equation (2) (reproduced above in (10)) attempts to provide a justification for their claims. It says

> The reformulation in Eq. (2) is justified by the assumptions above as to the identities of $A$ and $B$. Thus, $Pr[E|H_p,I_{a\backslash b},I_{b\backslash a},I_{ab}]$ may be written as $Pr[E|H_p,I_b]$ as $E$ is independent of $I_{a\backslash b}$ and $I_{b\backslash a} \cup I_{ab} = I_b$. Similarly, $Pr[E|H_d,I_{a\backslash b},I_{b\backslash a},I_{ab}] = Pr[E|H_d,I_b]$, $Pr[H_p|I_{a\backslash b},I_{b\backslash a},I_{ab}] = Pr[H_p|I_a]$, and $Pr[H_d|I_{a\backslash b},I_{b\backslash a},I_{ab}] = Pr[H_d|I_a]$.

In making the above assertions the authors of AN are saying "if $A$, $B$, $C$ are events such that $A$ and $B$ are independent, then $Pr[A|B,C] = Pr[A|C]$. This is a fallacious argument as the following simple example shows. Consider two independent tosses of a fair coin. Let $A$ be the event that the first toss results in 'heads', $B$ the event that the second toss results in 'heads' and $C$ the event that the outcomes of the two tosses are the same, that is, both are 'heads' or both are 'tails'. Then $A$ and $B$ are clearly independent but $Pr[A|B,C] = 1$ whereas $Pr[A|C] = 1/2$. Thus $Pr[A|B,C] \neq Pr[A|C]$. See Section 3.6, page 18 of (Stoyanov, 2013) for additional examples and discussions.

It is worth noting that the independence assumptions stated by the authors in AN (page 648, Column 2, lines 3-8) are different from the independence assumptions they invoke following their equation (2). In particular, they initially assumed that $I_a$ is formally independent of $E$ but, following their equation (2), they restate it as $I_{a\backslash b}$ is independent of $E$. However, these two versions of independence assumptions are not equivalent and neither is sufficient to justify their claims.

The following four equality conditions are needed to justify the arguments in equation (2) of `AN`.

$$Pr[E|H_p, I_{a\backslash b}, I_{b\backslash a}, I_{ab}] = Pr[E|H_p, I_b] \qquad (11)$$

$$Pr[E|H_d, I_{a\backslash b}, I_{b\backslash a}, I_{ab}] = Pr[E|H_d, I_b] \qquad (12)$$

$$Pr[H_p|E, I_{a\backslash b}, I_{b\backslash a}, I_{ab}] = Pr[H_p|E, I_a] \qquad (13)$$

$$Pr[H_d|E, I_{a\backslash b}, I_{b\backslash a}, I_{ab}] = Pr[H_d|E, I_a] \qquad (14)$$

The assumption needed to guarantee that (11) holds is that $E$ is conditionally independent of $I_{a\backslash b}$ given $H_p$ and $I_b$. But this is just a verbal description of what (11) says. Similarly, the assumption needed to guarantee that (12) holds is that $E$ is conditionally independent of $I_{a\backslash b}$ given $H_d$ and $I_b$ which is just a verbal description of what (12) is. Likewise, the assumption needed to guarantee that (13) and (14) hold is that $H_p$ is conditionally independent of $I_{b\backslash a}$ given $E$ and $I_a$.

In other words, the assumptions stated in `AN` do not lead to the stated result. To ensure (11) - (14) are true, one could directly assume them, but this would amount to the statement that "the hybrid modification to Bayes' equation holds if we make the simplifying assumption that it holds."

Given the failure to recognize the subjective elements of probability evaluations and the errors in attempted derivations, we do not consider `AN` to have refuted any of the concerns expressed in `LI`. Rather, our concerns are further strengthened by each instance in which proponents of experts communicating using $LR$s or the hybrid modification of Bayes' rule cite `AN` as having effectively addressed or dismissed any of the concerns expressed in `LI`.

## Aitken, Nordgaard, Taroni, Biedermann 2018

These authors begin their commentary with the following sentences.

> The first aspect concerns two related statements. In the abstract the statement is made that "[W]e find the likelihood ratio paradigm to be unsupported by arguments of Bayesian decision theory, which applies only to personal decision making and not to the transport of information from an expert to a separate decision maker." The idea presented in this statement of lack of support for the likelihood ratio as a means of transport of information is repeated in the conclusion where it is stated that "... we hope the forensic science community comes to view the LR as one possible, not normative or necessarily optimum, tool for communicating to DMs (decision makers)" (Lund and Iyer's emphasis). Despite this opinion of these authors, it was shown many years ago by I.J.Good in two brief notes in the Journal of Statistical Computation and Simulation (Good, 1989a,b) repeated in Good (1991) and in Aitken and Taroni (2004) that, with some very reasonable assumptions, the assessment of uncertainty inherent in the evaluation of evidence leads inevitably to the likelihood ratio as the only way in which this can be done.

It appears to us that these authors have misunderstood the concerns expressed in `LI`. In retrospect, we see how the title "Likelihood Ratios as Weight of Evidence: a Closer Look" might give one the impression that we would argue in favor of some probabilistic expression other than a likelihood ratio being a more appropriate characterization of an individual's perceived evidential strength.[18] This is not the case, however, as the concerns expressed in

---

[18] An additional point was made in `ANTB` that we had, incorrectly, used the term 'weight of evidence' in

LI were related to the practice of the forensic expert *communicating* to the decision maker their evidential findings through their own $LR$ assessment, $LR_B$. The first quote extracted from LI expressed concern with the use of the hybrid modification of Bayes' rule. The second extracted quote suggests that $LR$s are not the only way for experts to *communicate* with decision makers.

ANTB fail to address either of the extracted quotes used to motivate their rebuttal. They neither justify the hybrid modification to Bayes' rule nor demonstrate that it is normative or optimal for experts to *communicate* by providing their $LR$ to others. While AN attempts to refute the first point (which was addressed in the previous section), ANTB do not include any related arguments. To refute the second point, one would have to argue that it is normative or optimal for experts to *communicate* by providing their $LR$ to others. We are not aware of anyone even attempting to prove this theoretically or empirically. Rather, the content of an expert's report or testimony is simply a choice the expert (or the creator of their lab protocols) makes based on what s/he believes to be, but has not been demonstrated to be, "best practice."

Instead of addressing either point raised in the extracted quotes, ANTB appears to argue that (with some assumptions) *personal evaluation of evidence* necessarily leads to the $LR$. This is something of a straw man argument itself. As discussed in LI and the main body of this current article, we are in total agreement regarding the importance of $LR_A$ in updating $A$'s beliefs upon receiving new information $E_A$, and this was never disputed in LI. The distinction in LI is the argument that recipients must form their *own LR*s to comply with Bayesian reasoning (rather than using a value provided by an expert). LI did not suggest that recipients do something other than form an LR in response to provided information.[19]

---

reference to $LR_A$ and that we should be using the term 'value of evidence' to refer to it. The term "weight of evidence" is reserved for $log(LR_A)$. We agree and thank the authors of ANTB for pointing this out to us.

[19]LI discusses, in their sections 5.1 and 5.2, the role of likelihood ratios, both in the context of the Neyman-Pearson Lemma and in a Bayesian context. LI also explicitly displayed two forms of Bayes' equation, one that applies to the decision maker (DM) and one that applies to the expert.

Aside from failing to address the points they raise, we believe `ANTB` do not accurately represent the viewpoints of I. J. Good when paraphrasing his arguments (Good, 1989a; Good, 1989b; Good, 1991).

Here is what I. J. Good said in (Good, 1991).

> It will be convenient to imagine a trial by a single judge or magistrate without a jury. (Or we could imagine the thinking of a detective.) This will enable us to avoid complicating the issue by discussing multisubjective probabilities, although that is a worthwhile topic. (See, for example, DeGroot 1988.) Let us assume further that the judge is rational in the sense that he behaves as if he accepts the usual axioms of subjective (personal) probability. For a discussion of these axioms see, for example, Chapter 1 of this book or Good (1950, 1982a, b, 1983b, 1987). In the present chapter the main axiom is the product axiom, $P(C\&D) = P(C)P(D|C)$ which is a shorthand for $P(C\&D|F) = P(C|F)P(D|C\&F)$, where $C$, $D$, and $F$ are propositions asserting that various events occurred. Note that the vertical stroke, which is a standard notation (not an oblique stroke), always separates what occurs to its left and right, so it is unnecessary, for example, to write $(C\&D)$ in place of $C\&D$. If preferred, we can think of $C$, $D$, and $F$ as denoting events instead of propositions, where event is to be understood in a wide sense. Let us denote by $E$ the evidence presented in court, and by $I$ all the background knowledge or information available to the judge. We wish to express, in terms of probability, the weight of evidence in favour of guilt $G$ provided by the evidence $E$ given the background information $I$ all along. (The probabilities will usually be regarded as subjective or personal.) Denote this weight of evidence by $W(G : E|I)$, where the colon is read provided by and the vertical stroke by given. Thus $W$ depends upon three propositions.
>
> So far, we have merely introduced some notation, but now we make a critical assumption, namely that $W(G : E|I)$ can depend only on the probability of $E$ given that the man is guilty and on the probability given that he is innocent (really innocent, not merely 'found not guilty'). To state this condition more precisely, and in symbols, we assume that $W(G : E|I)$ depends only on $P(E|G\&I)$ and $P(E|G^c\&I)$. This assumption is, I believe, made throughout this book. It seems to us to be no more than common sense, but there are still some philosophers who suggest interpretations of $W(G : E|I)$ inconsistent with this assumption. At any rate, based on this piece of common sense, it can be proved, as a theorem, that $W(G : E|I)$ depends only on the ratio $P(E|G\&I)/P(E|G^c\&I)$. Various proofs of this simple important theorem, and allied theorems, have been been given (Good 1968, 1984, 1989a, b, c). We give the simplest proof (Good 1989c) in an appendix to this chapter, a proof that does not require any mention of the probabilities of guilt or innocence, and can therefore sometimes be interpreted as a 'non-Bayesian' proof. ...

We note the following points:

(a) Good begins by specifying that he is considering a single individual and states that this is to "avoid complicating the issue by discussing multisubjective probabilities, although that is a worthwhile topic." Hence, there is no consideration of communication between multiple parties other than to explicitly acknowledge that it is a different, more complicated, and important scenario to consider. Good recognizes the clear distinction between personal uncertainty evaluation and communication between two or more parties. LI and the current paper do so as well. ANTB does not.

(b) In the above paragraphs, Good is considering the issue of weighing the evidence $E$ from the perspective of a judge after receiving $E$ ("the evidence presented in court"). Notice also that Good is articulating how a rational judge would behave and how s/he might go about assessing the weight of the evidence provided. The hybrid modification of Bayes' rule (which the authors of ANTB continue to advocate) is clearly inconsistent with Good's representation of a recipient formulating their own weight of evidence in response to presented evidence.

(c) Good recognizes that the requirement (emphases are ours)

> *any quantity that might qualify as a 'weight of evidence' metric must be a function of $P(E|G\&I)$ and $P(E|G^c\&I)$*

as a *critical assumption*. This, as Good suggests, seems to be no more than common sense. But a key point that is overlooked in ANTB is that these probabilities are, as is clearly stated in Good's writing, those of the judge and not those of the expert.

(d) Let $x = P(E|G, I)$ and $y = P(E|G^c, I)$. Good first assumes that any quantity that is accepted as quantifying the weight of evidence must be a function of $x$ and $y$ alone, say $f(x, y)$. Furthermore, he considers an event or a proposition $F$ that is entirely irrelevant to $G$ and $E$, which, when written symbolically, is the statement that $P(G\&E|F, I) = P(G\&E|I)$.[20] He further observes that the weight of evidence pertaining to $G$ provided

---

[20]In paraphrasing Good's derivation, ANTB appears to have made an error. ANTB says "T may be taken to

by $E$ and $F$ together is the same as the weight of evidence provided by $E$ alone since $F$ is assumed to be entirely irrelevant to $G$ and $E$. This leads him to the consequence that weight of evidence has to be a function of the ratio $x/y$.

The choice of $Log(x/y)$ as a way to quantify weight of evidence is based on intuitive appeal of the additivity property when one visualizes each piece of evidence as some weight added to one side or the other side of the scales of justice. Good's derivation only says that a weight of evidence measure has to be a function of $LR$ if one is prepared to accept certain (reasonable) assumptions.

In conclusion, the comments in `ANTB` does not address any of the concerns expressed in `LI`.

---

be independent of E, of $H_p$, and of $H_d$" ($T$ in `ANTB` takes the place of $F$ in Good's derivation) rather than saying "$T$ may be taken to be jointly independent of $(E, H_p)$ (as Good implied). The pairwise independence that `ANTB` assumes is insufficient to justify the rest of the derivation in `ANTB`.